\documentstyle[11pt,seceq]{article}

\newcommand{\be}{\begin{equation}}
\newcommand{\ee}{\end{equation}}
\newcommand{\lb}{\label}

\newcommand{\bF}{{\bf f}}

\newcommand{\bh}{{\bf h}}
\newcommand{\bk}{{\bf k}}
\newcommand{\bm}{{\bf m}}
\newcommand{\br}{{\bf r}}

\newcommand{\bx}{{\bf x}}
\newcommand{\by}{{\bf y}}
\newcommand{\bz}{{\bf z}}
\newcommand{\bA}{{\bf A}}
\newcommand{\bB}{{\bf B}}
\newcommand{\bC}{{\bf C}}
\newcommand{\bD}{{\bf D}}
\newcommand{\bM}{{\bf M}}
\newcommand{\bQ}{{\bf Q}}
\newcommand{\bR}{{\bf R}}
\newcommand{\bV}{{\bf V}}
\newcommand{\bW}{{\bf W}}
\newcommand{\bX}{{\bf X}}
\newcommand{\bY}{{\bf Y}}
\newcommand{\bZ}{{\bf Z}}
\newcommand{\cA}{{\cal A}}
\newcommand{\cB}{{\cal B}}
\newcommand{\cC}{{\cal C}}
\newcommand{\cN}{{\cal N}}
\newcommand{\cP}{{\cal P}}
\newcommand{\cQ}{{\cal Q}}
\newcommand{\cR}{{\cal R}}
\newcommand{\cW}{{\cal W}}
\newcommand{\hL}{\hat{L}}
\newcommand{\obz}{\overline{{\bf z}}}
\newcommand{\boalpha}{{\mbox{\boldmath $\alpha$}}}
\newcommand{\bobeta}{{\mbox{\boldmath $\beta$}}}
\newcommand{\borho}{{\mbox{\boldmath $\rho$}}}
\newcommand{\bolambda}{{\mbox{\boldmath $\lambda$}}}
\newcommand{\bogamma}{{\mbox{\boldmath $\gamma$}}}
\newcommand{\bomu}{{\mbox{\boldmath $\mu$}}}
\newcommand{\boeta}{{\mbox{\boldmath $\eta$}}}
\newcommand{\bozeta}{{\mbox{\boldmath $\zeta$}}}
\newcommand{\boGamma}{{\mbox{\boldmath $\Gamma$}}}
\newcommand{\boZeta}{{\mbox{\boldmath $\Sigma$}}}
\newcommand{\boxi}{{\mbox{\boldmath $\xi$}}}
\newcommand{\bcZ}{{\mbox{\boldmath $\cal{Z}$}}}
\newcommand{\grad}{{\mbox{\boldmath $\nabla$}}}
\newcommand{\bdot}{{\mbox{\boldmath $\cdot$}}}
\newcommand{\bdots}{{\mbox{\boldmath $:$}}}
\newcommand{\bzed}{{\mbox{\boldmath $0$}}}

\begin{document}
\title{A Variational Formulation of Optimal Nonlinear Estimation}
\author{Gregory L. Eyink\footnote{Permanent address:
{\em Department of Mathematics,
University of Arizona, Tucson, AZ 85721}}
\\{\em CCS-3 MS-B256}\\
{\em Los Alamos National Laboratory}\\
{\em Los Alamos, NM 87545}}
\date{ }
\maketitle
\begin{abstract}
We propose a variational method to solve all three estimation problems for
nonlinear stochastic dynamical systems:
prediction, filtering, and smoothing. Our new approach is based upon a proper
choice of cost function, termed the
{\it effective action}. We show that this functional of time-histories is the
unique statistically well-founded cost
function to determine most probable histories within empirical ensembles. The
ensemble dispersion about
the sample mean history can also be obtained from the Hessian of the cost
function. We show that the
effective action can be calculated by a variational prescription, which
generalizes the ``sweep method''
used in optimal linear estimation. An iterative numerical scheme results which
converges globally to the variational estimator.
This scheme involves integrating forward in time a ``perturbed'' Fokker-Planck
equation, very closely related to the
Kushner-Stratonovich equation for optimal filtering, and an adjoint equation
backward in time, similarly related to the Pardoux-Kushner
equation for optimal smoothing. The variational estimator enjoys a somewhat
weaker property, which we call ``mean optimality''.
However, the variational scheme has the principal advantage---crucial for
practical applications---that it admits a wide variety of
finite-dimensional moment-closure approximations. The moment approximations are
derived reductively from the Euler-Lagrange
variational formulation and preserve the good structural properties of the
optimal estimator.
\end{abstract}

\section{Introduction}

The three classical problems of stochastic estimation are prediction,
filtering, and smoothing
of time series; e.g. see \cite{Gelb}. These correspond to estimating the
future, present, and
past states, respectively, based upon current available information. In more
detail, the nonlinear
estimation problem may be described as follows: assume as known some nonlinear
(Ito) stochastic
differential equation for a time-series $\bX(t)$:
\be d\bX = \bF(\bX,t)dt + (2\bD)^{1/2}(\bX,t)d\bW(t). \lb{1} \ee
Here $\bF$ is a (drift) dynamical vector, $\bD$ is a nonnegative diffusion
matrix, and
$\bW(t)$ is a vector Wiener process. Suppose also that some imperfect
observations $\br(t)$
are taken of a function $\bcZ(\bX(t),t)$ of the basic process, including some
measurement errors
$\borho(t)$ with covariance $\bR(t)$:
\be \br(t)= \bcZ(\bX(t),t) + \borho(t). \lb{2} \ee
It will generally be assumed that the distribution of the measurement errors is
known as well.
For example, the errors may be assumed to be proportional to a white noise:
$\borho(t)
= \bR^{1/2}(t)\boeta(t)$. Then the problem is, given the data ${\cal R}(t_f)=\{
\br(t): t<t_f\}$ up to
final measurement time $t_f$, to obtain the best estimate of $\bX(t)$ at times
$t>t_f,\,\,\,t=t_f$ and $t<t_f$.

The optimal filtering problem in the above general setting has been exactly
solved by Stratonovich
\cite{Strat} and Kushner \cite{Kush1,Kush2} within a Bayesian formulation.
Those authors have shown
that the conditional probability density ${\cal P}(\bx,t|{\cal R}(t))$, given
the current data
${\cal R}(t)$, obeys a stochastic partial differential equations, nowadays
called the
{\it Kushner-Stratonovich equation}. Explicitly, denoting ${\cal
P}_*(\bx,t)={\cal P}(\bx,t|{\cal R}(t)),$
the KS equation is of the form
\be \partial_t {\cal P}_*(\bx,t) = \hat{L}(t){\cal P}_*(\bx,t)
    + \bh^\top(t)[\bcZ(\bx,t)-\langle\bcZ(t)\rangle_{*t}]{\cal P}_*(\bx,t).
\lb{3} \ee
where
\be \hat{L}(t)= -\grad_\bx\bdot[\bF(\bx,t)(\cdot)] + \grad_\bx\otimes\grad_\bx
                                                   \bdots[\bD(\bx,t)(\cdot)]
\lb{4} \ee
is the standard Fokker-Planck linear operator, and
\be \bh(t)= \bR^{-1}(t)[\br(t)-\langle\bcZ(t)\rangle_{* t}] \lb{5} \ee
is a random forcing term constructed from the particular realization of the
observation $\br(t)$ obtained
in a given sample run of the system. Note that
$\langle\cdot\rangle_{*t}=E(\cdot|{\cal R}(t))$ denotes
conditional average, i.e. the average with respect to the distribution ${\cal
P}(\bx,t|{\cal R}(t))$ itself.
Hence, the KS equation is nonlinear. The integration of this equation forward
in time, with sequential input
of the fresh observations $\br(t)$ as they become available, solves, in
principle, the filtering problem.
The prediction problem is then solved in theory, by integrating the standard
Fokker-Planck equation
with ${\cal P}(\bx,t_f|{\cal R}(t_f))$ as initial data to obtain ${\cal
P}(\bx,t|{\cal R}(t_f))$ for $t>t_f$.

The optimal smoothing problem has also been solved, in principle, by Kushner
\cite{Kush4} and Pardoux \cite{Pard}.
They have shown that ${\cal P}(\bx,t|{\cal R}(t_f))$ for $t<t_f$ can be written
as
\be {\cal P}(\bx,t|{\cal R}(t_f)) = \cA_*(\bx,t)\cP_*(\bx,t), \lb{5a} \ee
where $\cP_*(\bx,t)$ is as above and $\cA_*(\bx,t)$ solves the adjoint equation
\be \partial_t {\cal A}_*(\bx,t) + \hat{L}^*(t){\cal A}_*(\bx,t)
    + \bh^\top(t)[\bcZ(\bx,t)-\langle\bcZ(t)\rangle_{*t}]{\cal A}_*(\bx,t)=0.
\lb{5b} \ee
This equation, with the random forcing $\bh(t)$, must be interpreted as a
``backward stochastic equation''.
It is solved subject to the final condition $\cA(\bx,t_f)=1$.

In certain cases, these estimators reduce exactly to solving a finite number of
ODE's. For example, in the linear case,
where $\bF(\bx,t)= \bA(t)\bx,\,\,\,\bD(\bx,t)=\bD(t)$, and
$\bcZ(\bx,t)=\bB(t)\bx,$ the KS optimal filter
reduces exactly to the finite-dimensional Kalman-Bucy optimal linear filter
\cite{KalBuc}. This reduction
occurs in the linear case because the conditional PDF is known rigorously to be
multivariate Gaussian,
uniquely specified by its mean and covariance. The conditional mean
$E[\bX(t)|{\cal R}(t)]$ coincides
with the Kalman-Bucy filter estimate $\boxi(t)$ of the current state $\bX(t),$
which is determined by
the solution of a stochastic ODE with sequential input of the observations. The
covariance matrix
$\bC(t)=E[\bX(t)\bX^\top(t)|{\cal R}(t)]-\boxi(t)\boxi^\top(t)$ is obtained as
well from a linear Ricatti
equation integrated forward in time.

The Pardoux-Kushner smoother is also finite-dimensional for linear systems. In
fact, it coincides there with
an alternative {\it variational formulation} of the linear estimation problem.
The latter can be motivated most
naively from the idea of least-square-error estimation. That is, one may
introduce a weighted square-error
functional for the dynamics,
\be \Gamma_X[\bx]= {{1}\over{4}}\int_{t_i}^{t_f}
dt\,\,[\dot{\bx}-\bA(t)\bx]^\top
                 \bD^{-1}(t)[\dot{\bx}-\bA(t)\bx], \lb{6} \ee
which measures the ``cost'' for a history $\bx(t)$ to depart from the solution
of the linear, deterministic
dynamics $\dot{\bx}=\bA(t)\bx$. The integral is weighted by the ``error
covariance'' $\bD(t)$ which arises
from the random noise. A similar cost function may be introduced for the
observation error of the data:
\be \Gamma_{R}[\borho]= {{1}\over{2}}\int_{t_i}^{t_f} dt\,\,\borho(t)^\top
\bR^{-1}(t)\borho(t). \lb{7} \ee
In that case, the solution to the estimation problem may be obtained by
minimizing with respect to $\bx$ the
combined cost function
\be \Gamma_{X,R}[\bx,\br]:= \Gamma_X[\bx] + \Gamma_{R}[\br-\bB\bx] \lb{8} \ee
when the set of observations $\{\br(t): t_i<t<t_f\}$ is input into the second
term. The minimizer $\bx_*=\bx_*[\br]$
is then the {\it optimal history}, which solves simultaneously all three
estimation problems. It
may be shown that $\bx_*(t)=\boxi(t)$ for $t\geq t_f$, so that the variational
estimator coincides
with the Kalman-Bucy filter and predictor. Furthermore, it may be shown that,
for $t<t_f$,
the variational estimator is given by the {\it Ansatz}
\be \bx_*(t) = \boxi(t) + \bC(t)\boalpha(t). \lb{9} \ee
Here, $\boalpha(t)$ is the solution of a linear {\it adjoint equation}
integrated backward in time
with the final condition $\boalpha(t_f)=\bzed$ and thus vanishes identically
for $t\geq t_f$.
However, it makes a contribution for $t<t_f$ to the smoother, proportional to
$\bC(t).$
This adjoint algorithm to calculate the minimizer is called the ``sweep
method'' \cite{Med},
and it gives the same result $\bx_*(t)= E\left[\bX(t)|\cR(t_f)\right]$ as
calculated by the Pardoux-Kushner equation.
Hence, it provides a finite-dimensional representation of the optimal smoother
for linear systems.

In general, however, the optimal estimators are infinite-dimensional, i.e. they
require the solution of (stochastic) PDE's.
For many of the spatially-extended, continuum systems of greatest interest in
geophysics and in engineering, this is, in fact,
a {\it functional} PDE. Even discretization for numerical solution results in a
(stochastic) PDE on a phase space of dimension literally
a billion or more. It has therefore been clear since their original formulation
that, for such spatially-extended or distributed systems,
the exact calculation of the optimal nonlinear estimator will be numerically
unfeasible. Kushner himself wrote an early paper \cite{Kush3},
in which he stressed this point and set up a formalism for approximating the
optimal filter. As he observed there, the problem is
formally the same as the ``closure problem'' in turbulence theory. The
approximation scheme he proposed was also the same as that
traditionally adopted in turbulence theory: namely, a {\it moment closure} of
the full KS equation. Such a scheme results in a
set of equations with a number of variables comparable to that in the starting
equation (\ref{1}), which may still be large but tractable.
Constructing finite-dimensional approximate estimators continues to be a
pressing research problem up to the present day, e.g. see \cite{BHL}.
Indeed, general approximation schemes for the full estimation problem
(prediction, filtering and smoothing) that are at once computationally
practicable and faithful to the optimal solution remain to be developed. The
fact that the problem of estimation for extended systems is
formally equivalent to the turbulence problem---a notoriously difficult
one---suggests that the solution here, too, will be nontrivial.
Not only must the formal properties of the optimal estimator be retained by any
approximation, but also the physical properties
of the underlying dynamical system must be sufficiently represented. The
problem of approximating the optimal estimator is not, in our opinion,
just one of mathematics but also of physics.

The aim of this paper is to formulate a new approach to the problem of optimal
nonlinear estimation, based upon a variational formulation.
The crux of the method is to identify an action functional, analogous to
(\ref{6}), which is statistically justified to use as a cost function
for estimation of nonlinear dynamics. This is the quantity which we have called
the {\it effective action} in previous works \cite{Ey1}-\cite{Ey3}.
This functional of state histories is uniquely characterized as that which
selects the most probable value under arbitrary conditions on the empirical
sample averages. In fact, the cost function (\ref{6}) appropriate for linear
systems has been motivated only rather crudely but it has a more fundamental
probabilistic justification. It was apparently first observed by the chemist
Lars Onsager that the dynamical cost function ({\ref6}) is the unique
functional whose minimum determines the statistically most probable
time-history of a {\it linear} dynamics of form (\ref{1}), subject to an
arbitrary
sets of constraints. In the statistical physics literature, the functional
(\ref{6}) is known as the {\it Onsager-Machlup action} \cite{OM}.
The cost function (\ref{7}) for the current observation error can be similarly
shown to give the most probable error in the case of a Gaussian
white-noise distribution. The combined cost function (\ref{8})---under the
assumption that dynamical noise and observation error are independent
random functions---then indeed gives by minimization the most probable
time-history subject to currently available information.
Thus, the minimizer $\bx_*[\br]$ is the unequivocal optimal estimator in the
linear case. Previous attempts to develop variational methods
for optimal nonlinear estimation have not paid sufficient attention to the
statistical requirements on the cost function.
For example, a functional has often been employed similar to (\ref{6}) for the
linear case,
\be \Gamma_X[\bx]= {{1}\over{4}}\int_{t_i}^{t_f}
dt\,\,[\dot{\bx}-\bF(\bx,t)]^\top
                 \bD^{-1}(t)[\dot{\bx}-\bF(\bx,t)], \lb{6a} \ee
naively based upon least-square-error philosophy. However, the use of this cost
function has no statistical justification, {\it except} in the
weak-noise limit ${\bf D}\rightarrow \bzed$. In that case, (\ref{6a}) is known
as the ``nonlinear Onsager-Machlup action'' and it is proved
to give the leading-order asymptotics of probabilities of time histories for
small noise \cite{Gr,FW}. However, except for the weak-noise limit
or for linear dynamics, the Onsager-Machlup action has no probabilistic
significance. Only the effective action---and no other cost function,
such as (\ref{6a}) above---will even have as its minimum the correct mean
value. The effective action thus plays the role of a ``fluctuation potential''
in the theory of empirical ensemble averages constructed from independent
samples, analogous to the Onsager-Machlup action for weak-noise
or for linear systems. In fact, the effective action is known to coincide with
the Onsager-Machlup action for weak-noise or for linear
systems \cite{Gr}. Thus, the optimal estimator proposed in this work coincides
with the standard ones for those special cases.

Unlike (\ref{6}), the effective action cannot generally be written as an
explicit function of the state histories. However, it has been shown
in \cite{Ey1}-\cite{Ey3} that it may be calculated by a constrained variational
method. The Euler-Lagrange equations that result are a
pair of forward and backward equations, very similar to the
Kushner-Stratonovich-Pardoux (KSP) equations. Despite this, the variational
estimator is not quite equivalent to the optimal estimator which follows from
the KSP equations. It possesses instead a property that we
call {\it mean-optimality}. Although somewhat weaker than the optimality
enjoyed by KSP, mean-optimality distinguishes it from other
``suboptimal'' estimators which have in fact no optimality whatsoever. However,
the main advance of the variational approach is in the problem
of constructing finite-dimensional approximations. Because the variational
estimator is based upon an Euler-Lagrange variational principle,
it is very easy to develop consistent approximations by a Rayleigh-Ritz scheme.
In this method, parameterized trial functions are selected to
represent the solutions of the forward-backward equations. Inserted into the
variational functional and varying over parameters, one obtains
approximations to the exact forward-backward equations and thereby to the
effective action. A straightforward use of this scheme in fact leads
to a moment-closure approximation for the forward filtering equation, much like
that originally proposed by Kushner \cite{Kush3}. However,
now also backward equations are obtained for the smoothing problem. Necessary
consistency properties with the forward equations
are guaranteed by the fact that these arise together as the Euler-Lagrange
system of an approximate action functional.

This paper is organized as follows: in Part I, we present our variational
formulation of optimal estimation. We first explain the
unique statistical significance of the effective action, which makes it
appropriate for variational estimation. We review there also
the definition and properties of the effective action, including the notions of
joint and conditional effective actions. The optimality property
will be established for the variational estimator and compared with that of the
KSP estimator. We next discuss how to calculate
the effective action based upon its variational characterization. An iterative
numerical scheme is outlined to numerically calculate
the variational estimator, which reduces to solving KSP-type equations. Some
matters important for practical
applications will finally be discussed: the case when measurements are taken,
not continuously, but at a discrete set of times, and the evaluation
of the ensemble dispersion around the most probable value of the sample mean.
In Part II the very important issue is addressed of constructing
finite-dimensional approximations to the variational estimator, crucial for
application of the methods to spatially-extended
(or distributed) systems with many degrees of freedom. A Rayleigh-Ritz
moment-closure scheme is developed, based upon the finite-dimensional
reduction of the nonequilibrium action. The use of this approximation scheme
for solution of practical estimation problems is finally discussed.

\renewcommand{\thesection}{\Roman{section}}

\section{Variational Formulation of Optimal Estimation}

\noindent {\bf I.1. Ensemble Theory of Estimation}
% \subsection{Ensemble Theory of Estimation}

\noindent There are intrinsic limits to our ability to estimate, which can be
understood most simply from
an {\it ensemble} point of view. If the stochastic dynamics (\ref{1}) is run
many times with different realizations of the noise or,
even in the deterministic case $\bD(t)\equiv \bzed$, if the initial data are
selected randomly from some starting distribution $\cP^{(0)}$,
then the solutions will be generally quite distinct. Thus, in $N$ different
trials there will be $N$ different outcomes $\bX_1(t),...,\bX_N(t)$.
It is therefore not obviously very meaningful to give a single value $\bx_*(t)$
as an estimate of the state given some partial information $\cR$
(unless, of course, that information included the {\it exact} initial data or
realization of the random noise!) It is true that the average over samples
will converge to the mean in the ensemble conditioned on the available
information:
\be \lim_{N\rightarrow\infty}{{1}\over{N}}\sum_{n=1}^N \bX_n(t)= E[\bX(t)|\cR].
\lb{6y1} \ee
However, the individual sample points will show a scatter, possibly quite
large, about this mean value. A useful measure of this scatter
is the covariance matrix
\be \bC_\cR(t):= E[\delta\bX(t)\delta\bX^\top(t)|\cR] \lb{6y2} \ee
in the ensemble conditioned on $\cR$, where
$\delta\bX(t):=\bX(t)-E[\bX(t)|\cR]$. In particular, ${\rm Tr}\bC_\cR(t):=
E[\|\delta\bX(t)\|^2|\cR]$ gives
the mean square radius $\sigma^2_\cR(t)$ of scatter of the sample points around
the mean. The ensemble mean has the one virtue that it {\it minimizes}
this rms radius of scatter. In other words, if one took
$\delta\bX(t):=\bX(t)-\bx_*(t)$ for any other non-random estimator
$\bx_*(t)\neq E[\bX(t)|\cR]$, one would increase $\sigma^2_\cR(t)$. This is an
elementary fact of probability theory: for any random variable,
the expectation value is the unique deterministic estimator for which the
mean-square error is a minimum. This important property of the mean
value as a predictor---that it minimizes rms forecast error---has been
emphasized before by Leith in the field of climatology \cite{Leith}.
Of course, the above considerations show that one should have not only an
estimate of the state of a system, but also an estimate
of the reliability or certainty of that state. The covariance matrix
$\bC_\cR(t)$ is a good such measure. Any state within a few standard
deviations $\sigma_\cR(t)$ of the mean must be regarded as having a good degree
of probability to occur.

Such considerations are precisely those which justify the standard Bayesian
approach of Kushner-Stratonovich-Pardoux.
Granted the limitations implied above, one cannot do better than to give the
probability density $\cP(\bx,t|\cR)$ of the state variable
conditioned on the available information. The minimal requirement on a
variational approach to estimation is thus that it should give at
least the mean and covariance of such conditioned ensembles. This has motivated
us to consider a cost function, the ``effective action'',
which has a proper foundation in the theory of empirical ensembles. A brief
review of its definition and basic properties is here required.

\noindent {\bf I.2. Basic Theory of the Effective Action}
%\subsection{Basic Theory of the Effective Action}

\noindent The quantity which we have termed the effective action
\cite{Ey1}-\cite{Ey3} has appeared,
in various guises and by various names, in quantum field theory, in theory of
stochastic processes,
and in dynamical systems theory. We shall here just briefly recapitulate its
definition
and properties.

One interpretation of the effective action is as a generating functional for
multi-time
correlations. This is the way in which the functional is generally introduced
in field theory
\cite{IZ}. Consider any vector-valued random process $\bZ(t)$. Then, the {\it
cumulant
generating functional} $W_Z[\bh]$ is defined as
\be W_Z[\bh]= \log\langle \exp\left(\int_{t_i}^{t_f}
dt\,\,\bh^\top(t)\bZ(t)\right)\rangle.
\lb{6b} \ee
The $n$th-order multi-time cumulants of $\bZ(t)$ are obtained from $W_Z[\bh]$
by functional
differentiation with respect to the ``test history'' $\bh(t)$:
\be C_{i_1\cdots i_n}(t_1,...,t_n)=
   \left. {{\delta^n W_Z[\bh]}\over{\delta h_{i_1}(t_1)\cdots\delta
h_{i_n}(t_n)}}\right|_{\bh=\bzed}.
\lb{6c} \ee
It is not hard to check from its definition (\ref{6b}) that $W_Z[\bh]$ is a
convex functional of $\bh$.
The Legendre dual of this functional is defined to be the effective action of
$\bZ(t)$:
\be \Gamma_Z[\bz]= \max_{\bh}\{ <\bh,\bz>-W_Z[\bh]\}, \lb{6d} \ee
with $<\bh,\bz>:=\int dt\,\,\bh^\top(t)\bz(t)$. It is a generating functional
of so-called
{\it irreducible correlation functions} of $\bZ(t)$:
\be \Gamma_{i_1\cdots i_n}(t_1,...,t_n)=
   \left. {{\delta^n \Gamma_Z[\bz]}\over{\delta z_{i_1}(t_1)\cdots\delta
z_{i_n}(t_n)}}\right|_{\bz=\obz}.
\lb{6e} \ee
The functional derivatives here are evaluated at the mean history
$\obz(t):=\langle \bZ(t)\rangle$.
It is not hard to check from the definition (\ref{6d}) that $\Gamma_Z[\bz]$ is
a convex, nonnegative
functional with a unique global minimum (equal to zero) at the mean history
$\bz=\obz$.

The effective action has another important interpretation as the rate function
in the theory
of large deviations of empirical ensemble averages for time-series. See
\cite{Cramer} for the original,
so-called {\it Cram\'{e}r theory} of single real variables and \cite{BZ} for
the extension to general
vector spaces. This theory involves the empirical or sample mean:
\be \overline{\bZ}_N(t):= {{1}\over{N}}\sum_{n=1}^N \bZ_n(t), \lb{6f} \ee
where $\bZ_n(t)$ for $n=1,...,N$ are independent, identically distributed
realizations of the
random process $\bZ(t)$. The law of large numbers states that, in the limit of
number of samples $N$
going to infinity, $\overline{\bZ}_N(t)\rightarrow \obz(t)$. However, for
finite $N$, $\overline{\bZ}_N(t)$
is itself a random process with some probability of achieving a fluctuation
value $\bz(t)$ different
from the ensemble mean $\obz(t)$. The basic result of the Cram\'{e}r theory is
that this probability
decreases exponentially in the limit as $N\rightarrow\infty$:
\be P\left(\overline{\bZ}_N(t)\approx \bz(t):t_i<t< t_f\right)
     \sim \exp\left(-N\cdot \Gamma_Z[\bz]\right). \lb{6g} \ee
Thus, $\Gamma_Z[\bz]$ for $\bz\neq\obz$ gives the rate of decay of the
probability to observe
$\overline{\bZ}_N\approx \bz$. Since $\Gamma_Z[\bz]=0$ only for $\bz=\obz$, the
probability
to observe the empirical $N$-sample mean $\overline{\bZ}_N$ equal to anything
other than the
ensemble mean $\obz$ must go to zero as $N\rightarrow\infty$. Thus, the large
deviation result (\ref{6g})
includes, and generalizes, the usual law of large numbers. Furthermore, the
ensemble mean
is now also seen to be characterized by a variational principle of {\it least
effective
action}. That is, the most probable value of the sample mean for large $N$, or
$\bz=\obz$,
is just that which minimizes the effective action $\Gamma_Z[\bz]$. It is worth
emphasizing that
the effective action is the {\it unique} function possessing all of these
properties. This is a consequence
of a general theorem on uniqueness of rate functions for large deviations
\cite{V}.

There is one other general theorem of large deviation theory which will prove
important to us in the
sequel. This is the so-called {\it Contraction Principle}. Suppose that
$\bW_N(t)$ is a random
process which is defined as a continuous functional ${\cal {\bf W}}$ of an
empirical mean $\overline{\bZ}_N(t)$:
\be \bW_N(t):= {\cal {\bf W}}[t;\overline{\bZ}_N]. \lb{6h} \ee
Then, $\bW_N(t)$ also satisfies a large deviations principle with rate function
given by
\be \tilde{\Gamma}_W[{\bf w}]= \min_{\{\bz:{\cal {\bf W}}[\bz]={\bf w}\}}
\Gamma_Z[\bz]. \lb{6i} \ee
See \cite{V}. When the functional ${\cal {\bf W}}$ is linear, then obviously
$\tilde{\Gamma}_W[{\bf w}]=\Gamma_W[{\bf w}].$

The {\it joint effective action} $\Gamma_{X,Y}[\bx,\by]$ of a pair of random
time series
$\bX(t)$ and $\bY(t)$ can be defined most simply as the effective action
$\Gamma_Z[\bz]$
of the composite vector $\bZ(t):=(\bX(t),\bY(t))$. Of course, this notion
may be extended to a joint effective action $\Gamma_{X_1\cdots
X_n}[\bx_1,...,\bx_n]$
of an arbitrary number $n$ of variables $\bX_i(t),\,\,i=1,...,n$. It is a
simple application
of the Contraction Principle to see that elimination of one of the variables is
accomplished
by minimizing over its possible values. For example,
\be \Gamma_X[\bx]= \min_{\by} \Gamma_{X,Y}[\bx,\by] \lb{6j} \ee
recovers the effective action $\Gamma_X[\bx]$ of $\bX(t)$ alone.

The joint effective action of a pair of {\it independent} time series $\bX(t)$
and $\bY(t)$
is easily shown from the definition to be given just by the sum:
\be \Gamma_{X,Y}[\bx,\by]= \Gamma_X[\bx]+\Gamma_Y[\by]. \lb{6k} \ee
In general, for dependent time series, one may define the notion of a {\it
conditional effective
action} by means of
\be \Gamma_{X|Y}[\bx|\by]:= \Gamma_{X,Y}[\bx,\by]-\Gamma_Y[\by]. \lb{6l} \ee
Thus, when $\bX(t)$ and $\bY(t)$ are independent,
$\Gamma_{X|Y}[\bx|\by]=\Gamma_X[\bx]$.
The term ``conditional action'' is justified by the relation to large-$N$
asymptotics
of conditional probabilities for empirical averages:
\be P\left(\overline{\bX}_N \approx \bx|\overline{\bY}_N=\by\right)
     \sim \exp\left(-N\cdot \Gamma_{X|Y}[\bx|\by]\right). \lb{6m} \ee
Using the definition of the conditional probability, (\ref{6m}) is a simple
consequence
of the basic large deviation estimate (\ref{6g}). The conditional action is
also a
generating functional for irreducible multi-time correlation functions in the
conditioned
ensemble, in the limit $N\rightarrow\infty$.

\noindent {\bf I.3. Proposal for the Variational Estimator}
% \subsection{Proposal for the Variational Estimator}

\noindent These considerations motivate our following proposal: {\it we propose
to take as} {\bf optimal
estimator} $\bx_*[\br]$ {\it the minimizer of the conditional effective action}
$\Gamma_{X|R}[\bx|\br]$
{\it of the history} $\bx$ {\it given the current observation history}
$\{\br(t): t\in[t_i,t_f]\}$.
{}From our discussion of the effective action in the previous section, we can
infer the crucial
property of this estimator: it is the mean value within the subensemble in
which the
empirical $N$-sample average takes on the value $\br$, that is, the
sub-ensemble in which
\be \overline{\br}_N(t)=\br(t),\,\,\,t\in[t_i,t_f]. \lb{10} \ee
In fact, as discussed above, the conditional effective action is a variational
functional whose minimum
coincides with the subensemble mean history for an arbitrary set of constraints
on the empirical sample average.
Thus, the estimator $\bx_*[\br]$ is exactly of the form $E[\bX(t)|\cR]$
discussed above, with $\cR$ specified by
(\ref{10}). It is also possible to obtain from the conditional effective action
the error covariance $\bC_*[t|\br]$,
essentially by evaluating its Hessian matrix (see Section I.6). This is
precisely the ensemble dispersion
that would be observed via the spread of sample histories in an ensemble
forecasting scheme \cite{Leith},
if the ensemble considered were the one specified by the condition (\ref{10})
for the limit of large $N$.

Our proposal is clearly similar in spirit to the Bayesian formulation of
Kushner-Stratonovich-Pardoux. However,
they are distinct. The difference can best be understood by considering the
problem from an experimental
point of view. Suppose that a very large number $N$ of samples of the system
(1) are run, drawing initial
conditions randomly from the same distribution ${\cal P}_0$. Then the
conditional distribution
${\cal P}(\bx,t|{\cal R}(t_f))$ considered by KSP corresponds to the very small
sub-ensemble
in which that particular realization $\br$ (non-random) of the observation
history occurred. It obviously
difficult to prepare such sub-ensembles, since one must wait patiently for the
particular observation
$\br$ to spontaneously occur and, each time it does, add it as a member to the
sub-ensemble. This makes
it very difficult to directly test the predictions of the KSP-equations. More
to the point, it is very difficult
to carry out an ensemble or Monte Carlo approach to calculate directly the
conditional average. The variational
estimation method proposed above corresponds to a different---and somewhat
larger---sub-ensemble. As noted above,
it corresponds to considering the sub-ensemble specified by the condition
$\{\overline{\br}_N(t)=\br(t),\,\,\,
t\in [t_i,t_f]\}$. It is clear that the sub-ensemble described by the
KSP-equations is, also, a subset of the new,
larger one. In fact, if it is true as in the KSP sub-ensemble that $\br_n=\br$
in {\it every} realization,
$n=1,...,N$ then it is {\it a fortiori} true that $\overline{\br}_N=\br$.
However, there will clearly be
many members of the new ensemble in which $\overline{\br}_N=\br$ but for which
not every term $\br_n$
of the $N$-sample average is equal to $\br$. Thus, the new sub-ensemble is
clearly much larger than that
considered by Kushner-Stratonovich-Pardoux, but, still, too small a subset of
the whole ensemble to be reproducible
by direct methods.

Despite the difficulty of directly testing the KSP-equations, they are truly
the optimal method for
the filtering and smoothing problems. While it is difficult to prepare the
conditional ensemble
corresponding to a given observation $\br$, there is no difficulty in preparing
{\it one member} of
such an ensemble. After all, just running the system once and collecting one
observation $\br$
provides one realization in which that particular observation occurs! In fact,
it is exactly
this type of situation which occurs in practical prediction problems, such a
meteorology. One has
no control over which particular weather pattern will be observed up to today,
but, given the
one that has occurred, one would like to predict tomorrow's weather. The best
predictor will be that
corresponding to the conditional ensemble in which all of the available
information is used.
The variational predictor we have proposed corresponds to a larger
sub-ensemble, which
means that somewhat less detailed information about the system is used in
making the prediction.
Therefore, the variational predictor is optimal, but in a somewhat
weaker sense than the KSP one. {\it The variational estimator} $\bx_*[\br]$
{\it is optimal given the data just on empirical sample
averages, which is somewhat less than the information one actually possesses.}
We shall refer to the weaker optimality
property possessed by the variational estimator as {\it mean-optimality}. In
the language of statistical physics,
the variational estimator could be termed a ``mean-field approximation'' to the
optimal one, since it exploits
conditions defined only through the sample mean. How different as predictors
are the variational
and KSP optimal estimators will depend upon how much variation occurs with the
larger sub-ensemble.
If all the terms $\br_n\approx \br$ in the sample average whenever
$\overline{\br}_N=\br$
for a given $\br$, then there will be little difference between the two
subensembles. This may
be expected to occur whenever there are ``preferred paths'' in the dynamical
evolution.

In the linear case, the variational method proposed above coincides with the
standard one described
in the Introduction. To see this is true it is enough to point out the
well-known fact that, for a
linear dynamics, the effective action $\Gamma_X[\bx]$ coincides with the
Onsager-Machlup action \cite{Gr}.
Hence, in the linear case, the variational estimator coincides with
the optimal estimator of Kushner-Stratonovich-Pardoux, so far as the
problems of prediction and filtering are concerned. This will not be true in
general for nonlinear systems.
As we shall see in the next subsection, there is nevertheless a close formal
connection between the Bayesian
approach of Kushner-Stratonovich-Pardoux and the variational approach.

\noindent {\bf I.4 Calculation of Effective Action \& Variational Estimator}
% \subsection{Calculation of Effective Action \& Variational Estimator}

\noindent It remains to consider how the effective action and its minimizer,
the variational estimator, may actually be calculated.
It was shown in \cite{Ey1,Ey3} that the effective action may be obtained from a
constrained variational formulation. We shall here
briefly review the results of those works and then explain how to obtain the
minimizing history $\bx_*[\br]$ itself.

Consider any Markov times series $\bX(t)$ and $\bZ(t):=\bcZ(\bX(t),t)$ given by
a
continuous function $\bcZ(\bx,t)$. In this general context there is a useful
variational characterization
of the effective action $\Gamma_Z[\bz]$. To explain this result, we must
introduce a few notations.
Because the process $\bX(t)$ is Markov, its distribution $\cP(\bx,t)$ at time
$t$ is governed by the
{\it forward Kolmogorov equation}
\be \partial_t \cP(\bx,t) = \hL(t)\cP(\bx,t), \lb{6n} \ee
with $\hL(t)$ the instantaneous Markov generator. The diffusion process
governed by the stochastic
equation (\ref{1}) is a particular example, for which the generator is the
Fokker-Planck operator
defined in (\ref{4}). Observables, or random variables, $\cA(\bx,t)$ evolve
under the corresponding
{\it backward Kolmogorov equation}
\be \partial_t \cA(\bx,t) = -\hL^*(t)\cA(\bx,t), \lb{6o} \ee
in which $\hL^*(t)$ is the adjoint operator of $\hL(t)$ with respect to the
canonical bilinear
form on $L^\infty\times L^1$, i.e. $<\cA,\cP>:= \int d\bx \,\,\cA(x)\cP(\bx)$.
The backward and
forward Kolmogorov equations may be simultaneously obtained as Euler-Lagrange
equations
for stationarity of the {\it action functional}
\be \Gamma[\cA,\cP]:= \int_{t_i}^{t_f} dt\,\,<\cA(t),(\partial_t-\hL(t))\cP(t)>
\lb{6p} \ee
when varied over $\cP\in L^1$ with initial condition $\cP(t_i)=\cP_0$ and
$\cA\in L^\infty$ with
final condition $\cA(t_f)\equiv 1$.

For the above situation, the effective action of $\bZ(t):=\bcZ(\bX(t),t)$ has
been shown \cite{Ey1,Ey3}
to be obtained by a {\it constrained variation} of the action
$\Gamma[\cA,\cP]$. In fact,
\be \Gamma_Z[\bz]= {\rm st.pt.}_{\cA,\cP} \Gamma[\cA,\cP] \lb{6q} \ee
when varied over the same classes as above, but subject to constraints of fixed
overlap
\be <\cA(t),\cP(t)>= 1 \lb{6r} \ee
and fixed expectation
\be <\cA(t),\hat{\bZ}(t)\cP(t)>= \bz(t) \lb{6s} \ee
for all $t\in [t_i,t_f]$. Note that $\hat{\bZ}(t)$ is used to denote the
operator (in both
$L^1$ and $L^\infty$ ) of multiplication by $\bZ(\bx,t)$. The Euler-Lagrange
equations for
this constrained variation may be obtained by incorporating the expectation
constraint (\ref{6s})
with a Lagrange multiplier $\bh(t)$. The overlap constraint could also be
imposed with
a Lagrange multiplier $w(t)$. However, it turns out to be preferable to impose
it through the
definitions
\begin{eqnarray}
\cA(t) & = & 1 +\left[\cB(t)-\langle\cB(t)\rangle_t\right] \cr
    \, & := & 1 + \cC(t), \lb{6t}
\end{eqnarray}
with the final conditions $\cB(t_f)=\cC(t_f)\equiv 0$. Note that
$\langle\cB(t)\rangle_t:= <\cB(t),\cP(t)>$
is the expectation with respect to the distribution $\cP(t)$. Hence, the
overlap constraint (\ref{6r})
is satisfied when $\cB(t)$ is varied independently of $\cP(t)$. Like $\cA(t)$,
the variable $\cC(t)$ is not
independent of $\cP(t)$, but must satisfy the orthogonality condition
$<\cC(t),\cP(t)>=0$. We shall mostly make use here of the original
variable $\cA(t)$ rather than $\cC(t)$, but the latter will play an important
role in our formulation of moment-closures in Part II.

Although obtained by varying over $\cB(t),\cP(t)$, the Euler-Lagrange equations
are most usefully
written instead in terms of the original variables $\cA(t),\cP(t)$:
\be \partial_t \cP(t) =
\hat{L}(t)\cP(t)+\bh^\top(t)[\bcZ(t)-\langle\bcZ(t)\rangle_t]\cP(t)
    \lb{6u} \ee
and
\be \partial_t \cA(t)+
\hat{L}^*(t)\cA(t)+\bh^\top(t)[\bcZ(t)-\langle\bcZ(t)\rangle_t]\cA(t)=0.
\lb{20r} \ee
% They can also be written in terms of $\cC(t),\cP(t)$, with equation
%%(\ref{6u}) unchanged and (\ref{20r}) replaced by
% \be \partial_t \cC(t)+
%%\hat{L}^*(t)\cC(t)+\bh^\top(t)[\bcZ(t)-\langle\bcZ(t)\rangle_t]\cC(t)
%                      +\bh^\top(t)[\bcZ(t)-\langle\bcZ(t)\rangle_t]= 0.
%    \lb{6v} \ee
% In terms of $\cC(t),\cP(t)$ the expectation constraint is
% \be \langle\bcZ(t)\cC(t)\rangle_t + \langle\bcZ(t)\rangle_t = \bz(t). \lb{6w}
%%\ee
% The effective action is obtained from the solutions of (\ref{6u}),(\ref{6v})
%%as
% \be \Gamma_Z[\bz]= \int_{t_i}^{t_f} dt\,\,<\cC(t),(\partial_t-\hL(t))\cP(t)>,
%%\lb{6x} \ee
% when the ``control field'' $\bh(t)$ is chosen so that (\ref{6w}) reproduces
%%the considered history
% $\bz(t)$.
The calculation via $\cB(t)$ has allowed the Lagrange multiplier to be
evaluated explicitly, as $w(t)=\bh^\top(t)\langle\bcZ(t)\rangle_t.$
The effective action $\Gamma_Z[\bz]$ evaluated at a specific history $\bz(t)$
is now obtained from the solutions of (\ref{6u}),(\ref{20r}) by
substituting them back into the action functional $\Gamma[\cA,\cP]$ in
(\ref{6p}), when the ``control field'' $\bh(t)$ is chosen so that
(\ref{6s}) reproduces the considered history $\bz(t)$. It is not accidental
that the same notation $\bh(t)$ was chosen above as for the argument
of the cumulant generating functional $W_Z[\bh]$. In fact, it can be shown that
also
\be W_Z[\bh]= \int_{t_i}^{t_f} dt\,\, \bh^\top(t)\langle\bcZ(t)\rangle_t,
\lb{6y} \ee
using just the solution $\cP(t)$ of the forward equation (\ref{6u}) for the
control history $\bh(t)$ which appears as the argument of $W_Z$.
For more details, see \cite{Ey1,Ey3}. It should not have escaped the attention
of the reader that the forward equation (\ref{6u})
is very similar to the Kushner-Stratonovich equation (\ref{3}) for the
conditional distribution $\cP_*(t)=\cP(t|\cR(t))$ and that the backward
equation (\ref{20r}) is likewise similar to the Kushner-Pardoux equation
(\ref{5b}) for $\cA_*(t)=\cP(t|\cR(t_f))/\cP(t|\cR(t))$.
This observation will be developed below. (See also Appendix 1.)

Having completed our review of established results, we now consider
how to calculate the variational estimator. It is helpful to observe that the
minimizer $\bx_*[\br]$
of $\Gamma_{X|R}[\bx|\br]$ over $\bx$ with $\br$ fixed is the same as of
$\Gamma_{X,R}[\bx,\br],$
the joint action of $\bx$ and $\br$. For simplicity, the observation errors
will be assumed to be
white-noise in time and independent of the dynamical noise. Another important
simplifying assumption
we shall make here is that the function of the process which is observed is
linear:
\be \bcZ(\bx,t)= \bB(t)\bx. \lb{11g} \ee
We postpone to later the consideration of the general case, which is somewhat
more complicated but no different
in principle. By our assumptions, the joint action is given as
\be \Gamma_{X,R}[\bx,\br]= \Gamma_{X}[\bx]
+ {{1}\over{2}}\int_{t_i}^{t_f} dt\,\,[\br(t)-\bB(t)\bx(t)]^\top
\bR^{-1}(t)[\br(t)-\bB(t)\bx(t)].  \lb{11} \ee
The second term is $\Gamma_R[\borho]$ given in (\ref{7}). We abbreviate
$\Gamma_*[\bx]:=\Gamma_{X,R}[\bx,\br]$
and its functional derivative as
$\bk_*[t;\bx]={{\delta\Gamma_*}\over{\delta\bx(t)}}[\bx]$. It is an easy
calculation,
using the expression (\ref{11}), to show that
\be \bk_*[t;\bx]= \bk[t;\bx]+
\bB^\top(t)\bR^{-1}(t)\left[\bB(t)\bx(t)-\br(t)\right], \lb{12} \ee
with $\bk[t;\bx]:={{\delta\Gamma_X}\over{\delta\bx(t)}}[\bx]$. Observe that we
are using here
the notation $\bk(t)$ for the control associated to $\bX(t)$, whereas we
reserve $\bh(t)$
for the control field associated to $\bZ(t)$. What makes finding the minimizer
$\bx_*[\br]$ less trivial
is the fact that $\Gamma_X[\bx]$ and $\bk[t;\bx]$ are not calculable directly,
but only as the result
of another optimization problem, like that in Eq.(\ref{6d}):
$\Gamma_X[\bx]=\max_\bk\left\{<\bx,\bk>-W_X[\bk]\right\}.$
Thus, the problem to be solved is really of {\it minimax} type:
\be \Gamma_*[\bx_*[\br]] = \min_\bx \max_\bk \left\{
\Gamma_R[\br-\bB\bx]+<\bx,\bk>-W_X[\bk]\right\}. \lb{11a} \ee
Numerical schemes to obtain the minimizer $\bx_*[\br]$ must thus address this
minimax problem.

The simplest approach conceptually is to reformulate it as a double
minimization, i.e.
\be \Gamma_*[\bx_*[\br]] = \min_\bx \left\{
\Gamma_R[\br-\bB\bx]-\min_\bk\{W_X[\bk]-<\bx,\bk>\}\right\}. \lb{11b} \ee
In this case, each of the minimizations may be carried out in nested fashion,
via any of the common iterative
methods. For example, a {\it conjugate gradient} (CG) algorithm applied to the
outer problem will produce a sequence
$\bx^{(n)}$ converging as $n\rightarrow\infty$ to, at least, a local minimum
$\bx_*$ of $\Gamma_*[\bx]$.
We mention conjugate gradient only as an example of an iterative scheme to find
the minimum of a convex function,
which requires as its input at each step the gradient
$\bk_*^{(n)}(t)={{\delta\Gamma_*}\over{\delta\bx(t)}}[\bx^{(n)}]$.
Any such scheme requiring the gradient might be used instead. From (\ref{12})
such algorithms require knowing
$\bk[\bx^{(n)}]$. Conveniently, this is exactly what is obtained from the
solution of the inner problem, since $\bk[\bx^{(n)}]$
is the unique minimizer $\bk^{(n)}$ of the convex functional
$W^{(n)}[\bk]:=W_X[\bk]-<\bk,\bx^{(n)}>.$ This inner
minimization problem may also be attacked by a CG-type method, noting that the
gradient is
\be {{\delta W^{(n)}}\over{\delta\bk(t)}}[\bk]= \bx[t;\bk]-\bx^{(n)}(t).
\lb{13} \ee
This gradient is now directly calculable via formula (\ref{6s}) above for a
given $\bk(t)$. Each evaluation
of $\bx[\bk]$ by (\ref{6s}) requires one forward and one backward integration
over the time interval $[t_i,t_f]$.
A CG-type method applied to $W^{(n)}[\bk]$ will then produce a sequence
$\bk^{(n,m)}$ which converges
to $\bk^{(n)}=\bk[\bx^{(n)}]$ as $m\rightarrow\infty$. This inner minimization
thus provides the gradient $\bk^{(n)}$
required for the $n$th CG step of the first minimization. To initiate the
algorithm, one must specify $\bx^{(0)}$
and $\bk^{(0,0)}$. For this purpose, one may, for example, set $\cA\equiv 1$ as
a first approximation in (\ref{6s}).
This gives
\be \bx^{(0)}(t)=\langle \bX(t)\rangle_t \lb{14} \ee
and, from the equation $\bk_*[t;\bx^{(0)}]= \bzed$, the first guess
\be \bk^{(0,0)}(t)=  \bB^\top (t)\bR^{-1}(t)\left[\br(t)-\bB(t)\bx(t)\right]
    \lb{15} \ee
If (\ref{15}) is substituted into the forward equation (\ref{6u}), the latter
may be integrated
with sequential input of the observations $\br(t)$. Thence, both $\bx^{(0)}$
and $\bk^{(0,0)}$
are determined. At each successive stage one may take $\bk^{(n+1,0)}=\bk^{(n)}$
to find
the gradient $\bk^{(n+1)}$ for the $(n+1)$st CG step. This entire procedure can
be regarded
as a nonlinear generalization of the ``sweep method'' \cite{Med} used to find
the minimizer
of the Onsager-Machup action (\ref{6}).

While this method has the advantage of conceptual simplicity, it suffers
numerically from loss of precision
and computational inefficiency. It is well-known in numerical optimization that
minimizers are in general obtained
to only half the precision of the minimum values themselves. As it is the
outside minimizer which is of
direct interest here, the double minimization algorithm requires working in a
precision {\it quadruple}
to that desired for the optimizing history. Furthermore, the nested algorithm
requires the {\it square}
of the number of iterations as for a single minimization. It is thus
advantageous to reformulate the minimax
problem in terms of a single numerical minimization. This can be easily
accomplished by rewriting it as
\be \Gamma_*[\bx_*[\br]] = \min_\bk \left\{
\Gamma_R[\br-\bB\bx[\bk])]+<\bx[\bk],\bk>-W_X[\bk]\right\}. \lb{11c} \ee
(We thank M. Anitescu for this observation.) Note again that $\bx[t;\bk]$ is
given directly by (\ref{6s}) via
one integration each of the forward and backward Kolmogorov equations over the
time interval $[t_i,t_f]$.
The result of this single minimization is a control field $\bk_*[\br]$, which
then yields the desired optimal history
$\bx_*[\br]$ as $\bx[\bk_*[\br]]$. The only disadvantage of this formulation is
that the gradient of the
functional in brackets in (\ref{11c}),
\be G_X[\bk,\br]:=  \Gamma_R[\br-\bB\bx[\bk]]+<\bx[\bk],\bk>-W_X[\bk], \lb{11d}
\ee
is
\be {{\delta G_X}\over{\delta \bk(t)}}[\bk,\br] = \int_{t_i}^{t_f} dt' \,\,
{{\delta\bx}\over{\delta\bk(t)}}[t';\bk]\left[\bk(t')+\bR^{-1}(t')
\left(\bB(t')\bx[t';\bk]-\br(t')\right)\right]. \lb{11e} \ee
This expression involves
\be {{\delta\bx}\over{\delta\bk(t)}}[t';\bk]={{\delta^2
W_X}\over{\delta\bk(t)\delta\bk(t')}}[\bk], \lb{11f} \ee
the Hessian of the dual functional $W_X[\bk]$. Thus, this 2nd-derivative must
be evaluated and stored for use.
The storage issue is nontrivial for spatially-extended or distributed systems,
because the Hessian then involves
a number of elements of the order of the spacetime grid squared. However, these
problems can be overcome.
First, there are efficient direct and adjoint algorithms for calculating
higher-order derivatives, such as Hessians,
in addition to those for first derivatives. For example, see \cite{NocWr},
Chapter 7, and also \cite{Ey3}. Second,
it is not really the Hessian itself which must be stored but only its matrix
products with certain vectors,
those in (\ref{11e}). Hence, storage requirements can be reduced in intelligent
schemes to vectors of the
same order as required for the double minimization algorithm. We give further
details of such algorithms
elsewhere, which we regard as the most promising numerical implementations of
our estimation method.

Whichever of these iterative optimization methods is employed, $\Gamma_*[\bx]$
is a convex functional, and the iterates
will therefore converge to the global minimizer $\bx_*[\br].$ Observe that the
zeroth-order of the double iteration scheme
coincides formally with the KSP equations (\ref{3})-(\ref{5b}). In fact, it is
then easy to see that equation (\ref{15})
for $\bk(t)$ at zeroth-order reduces to
\be \bk(t)= \bB^\top(t)\bh(t), \lb{11h} \ee
with $\bh(t)$ given precisely by (\ref{5}). Substituting this value, the
forward-backward equations in our
iterative scheme reduce in form to the KSP equations (\ref{3}),(\ref{5b}). In
general, there is no reason
to believe (except for linear dynamics), that the variational filter and KS
filter will coincide.
However, one may hope that the variational estimator, acting as a filter, is
not too far from the optimal
KS filter. The formal coincidence of these two in the case of linear
observations at the start
of the iterative construction provides possibly a convenient algorithmic
approach to assess
the differences. We emphasize, however, the word ``formal'' in this context,
because the variational equations (\ref{6u}), (\ref{20r}),
while appearing in form identical to the KSP equations (\ref{3}),(\ref{5b}),
have a quite different mathematical interpretation. Whereas
the control field $\bh(t)$ in the variational equations is non-random, the KSP
equations are stochastic PDE's. In particular, the numerical
discretization schemes appropriate to the two mathematical interpretations are
quite different and lead to quantitatively distinct
results. This will be discussed in more detail below for the case of
discrete-time measurements.

% \noindent {\bf Appendix 1: Variational Estimation with Nonlinear
%%Measurements}

When the measured function $\bcZ(\bx,t)$ is nonlinear in $\bx$, then our
approach must be slightly
generalized. In this case, we consider the joint action
$\Gamma_{X,Z,R}[\bx,\bz,\br]$, whose minimum over
$\bx,\bz$ with $\br$ fixed yields the optimum state estimate $\bx_*[\br]$ and
also the optimum value of the
measured variable $\bz_*[\br]$. The advantage to considering this joint action
is that it is simply expressed
in terms of the effective action $\Gamma_R[\borho]$ of the observation error,
which is still assumed independent
but not necessarily Gaussian. Indeed, a simple calculation in this case gives
\be \Gamma_{X,Z,R}[\bx,\bz,\br]= \Gamma_{X,Z}[\bx,\bz] + \Gamma_R[\br-\bz].
\lb{17b} \ee
In contrast, the joint action $\Gamma_{X,R}[\bx,\br]$ does not have such a
simple expression, but instead
must be calculated via the Contraction Principle as
$\Gamma_{X,R}[\bx,\br]=\min_\bz \Gamma_{X,Z,R}[\bx,\bz,\br].$
In the case of a linear observed variable, $\bcZ(\bx,t)=\bB(t)\bx$, the joint
action $\Gamma_{X,Z}[\bx,\bz]$ is found to be
\be \Gamma_{X,Z}[\bx,\bz]= \left\{ \begin{array}{ll}
                                    \Gamma_X[\bx] & {\rm if} \,\,\,\,
\bz=\bB\bx \cr
                                    +\infty   & {\rm otherwise}
                                   \end{array} \right. \lb{19f} \ee
Hence $\Gamma_{X,R}[\bx,\br]= \Gamma_X[\bx]+\Gamma_R[\br-\bB\bx]$ and the
estimation strategy we have proposed
for a nonlinear measurement function reduces to the earlier one in the linear
case.

The minimization of $\Gamma_{X,Z,R}[\bx,\bz,\br]$ over $\bx,\bz$ may be done in
two steps, which can be
carried out independently. These are, first, to minimize
\be \Gamma_{Z,R}[\bz,\br]= \Gamma_{Z}[\bz] + \Gamma_R[\br-\bz]. \lb{16} \ee
over all $\bz$ at fixed $\br$, and, second, to minimize $\Gamma_{X,Z}[\bx,\bz]$
over all $\bx$ with $\bz$ fixed.
{}From the solutions of these two problems, $\bz_*[\br]$ and $\bx_*[\bz]$,
respectively, the final variational estimator
of the state of the system is then the obtained as the composition
$\bx_*[\br]=\bx_*[\bz_*[\br]]$.
The equivalence of this two-step formulation with the direct one is an
application of the Contraction Principle.
Clearly, minimizing $\Gamma_{X,R}[\bx,\br]$ over all $\bx$ can be achieved by
minimizing first $\Gamma_{X,Z,R}[\bx,\bz,\br]$
over all $\bx$ with $\bz$ fixed, and then by minimizing over all $\bz$. The
minimization over $\bx$ yields the joint
effective action of $\bz$ and $\br$, since
$\Gamma_{Z,R}[\bz,\br]=\min_\bx\Gamma_{X,Z,R}[\bx,\bz,\br]$ by the Contraction
Principle.
The minimum is achieved here for some $\bx_*[\bz]$, the optimal state history
$\bx$ for a given $\bz$-history.
There is no dependence upon $\br$. To see this, observe that the minimization
may be directly carried out in equation (\ref{17b}),
with the result that $\Gamma_{Z,R}[\bz,\br]$ is given by (\ref{16}). The
Contraction Principle has been employed again
to infer $\Gamma_{Z}[\bz]=\min_\bx \Gamma_{X,Z}[\bx,\bz]$. It is from this
minimization that $\bx_*[\bz]$ is determined,
which therefore cannot involve $\br$. All of the dependence upon measurements
is now isolated in (\ref{16}),
whose minimization over $\bz$ yields $\bz_*[\br]$.

This first minimization of $\Gamma_{Z,R}[\bz,\br]$ over $\bz$ is a problem of
the same type as for the case of linear
measurement functions discussed in the text. As there, a CG-type method applied
to $\Gamma_*[\bz]:=\Gamma_{Z,R}[\bz,\br]$
may be employed to calculate $\bz_*[\br],$ based upon the Legendre dual
relations
\be \bh[t;\bz]={{\delta \Gamma_Z}\over{\delta\bz(t)}}[\bz],\,\,\,
    \bz[t;\bh]={{\delta W_Z}\over{\delta\bh(t)}}[\bh]. \lb{17} \ee
Any of the algorithms discussed in the text may be employed. For example, in
the double minimization scheme,
the gradient for the outer minimization,
\begin{eqnarray}
\bh_*[t;\bz^{(n)}] & := & {{\delta\Gamma_*}\over{\delta\bz(t)}}[\bz^{(n)}] \cr
          \, & =  & \bh[t;\bz^{(n)}]+ \bR^{-1}(t)[\bz^{(n)}(t)-\br(t)],
\lb{17a}
\end{eqnarray}
would be obtained from an inner one. The iteration could be initiated by
\be \bz^{(0)}(t)=\langle \bZ(t)\rangle_t \lb{18} \ee
and
\be \bh^{(0,0)}(t)= \bR^{-1}(t)\left[\br(t)-\langle \bZ(t)\rangle_t\right].
\lb{19} \ee
Just as before---but now quite in general---the zeroth-order control
$\bh^{(0,0)}(t)$, when substituted
into the forward-backward equations (\ref{6u}), (\ref{20r}) recovers formally
the KSP equations.

The second minimization of $\Gamma_{X,Z}[\bx,\bz]$ over $\bx$ is similar. Note
that
\be   \Gamma_{X,Z}[\bx,\bz]=
\max_{\bk,\bh}\left\{<\bk,\bx>+<\bh,\bz>-W_{X,Z}[\bk,\bh]\right\}. \lb{19c} \ee
Hence, the problem
\be   \Gamma_{X,Z}[\bx_*[\bz],\bz]= \min_\bx
\max_{\bk,\bh}\left\{<\bk,\bx>+<\bh,\bz>-W_{X,Z}[\bk,\bh]\right\}
                \lb{19d} \ee
is again of minimax type. A doubly iterative scheme would therefore carry out
the maximization over $\bk,\bh$
at fixed $\bx,\bz$ to obtain not only $\Gamma_{X,Z}[\bx,\bz]$ but also the
gradients
\be \bk[t;\bx,\bz]=
{{\delta\Gamma_{X,Z}}\over{\delta\bx(t)}}[\bx,\bz],
\,\,\,\bh[t;\bx,\bz]={{\delta\Gamma_{X,Z}}\over{\delta\bz(t)}}[\bx,\bz]
       \lb{19g} \ee
that are used in the next minimization over $\bx$ (at fixed $\bz$).
Alternatively, one may solve this problem
as before via a single minimization over $\bk,\bh$ of a functional
\be  G_{X,Z}[\bk,\bh]:= <\bk,\bx[\bk,\bh]>+<\bh,\bz[\bk,\bh]>-W_{X,Z}[\bk,\bh]
\lb{19e} \ee
but with the difference that this minimization is now subject to a nonlinear
constraint that
\be \bz[t;\bk,\bh]=\bz(t),\,\,\,\,\,\,t\in[t_i,t_f]. \lb{19h} \ee
This may be addressed using algorithms from nonlinear programming or
stochastic/ quasi-random methods.

\newpage

\noindent {\bf I.5. Estimation with Discrete-Time Data}
% \subsection{Estimation with Discrete-Time Data}

\noindent So far, we have considered the case where the measurements employed
in our estimation are
taken continuously in time. However, this can only be an idealization of a
situation where
the data are obtained at a discrete series of times. In many practical
examples, the instants
of measurement will be so widely separated that the idealization of continuous
acquisition
is far from valid. It is thus a very practical concern to address the issue of
state estimation
of continuous in time, stochastic dynamical systems such as (\ref{1}) based
upon discrete-time data.
In addition, we shall find that some fundamental new concepts are required that
are important in other
contexts. For example, the calculation of ensemble dispersions at an instant of
time will turn out to
be closely related to the problem of estimation with discrete-time data.

The only change in the statement of the problem in the Introduction is that now
the measurements
are of the form
\be \br_k = \bcZ(\bx(t_k),t_k) + \borho_k,\,\,\,k=1,...,n \lb{20a} \ee
where $\borho_k$ represents a measurement error with covariance $\bR_k$. If the
measurement error
is taken to be an independent Gaussian at each time $t_k$, then the cost
function for the observations is
\be \Gamma_{R}[\borho] = {{1}\over{2}}\sum_{k=1}^n \borho_k^\top \bR_k^{-1}
\borho_k, \lb{20b} \ee
where the sum includes all of the observation times $t_1,...,t_n$ up to the
present time. The combined
cost function $\Gamma_*[\bz]:=\Gamma_{Z,R}[\bz,\br]$ for the estimation is
then, analogous to (\ref{16}),
\be \Gamma_*[\bz]= \Gamma_{Z}[\bz]
             + {{1}\over{2}}\sum_{k=1}^n [\br_k-\bz(t_k)]^\top \bR_k^{-1}
[\br_k-\bz(t_k)]. \lb{20c} \ee
For simplicity, we shall only consider here the problem of estimating the
optimal $\bz$-history. As
discussed in the previous section, there remains the problem of estimating the
optimal state or $\bx$-history,
given the $\bz$-history. This can be handled in the same way as discussed
there. Alternatively,
we might formulate the problem as a direct estimation of $\bx$. The changes
necessary to our discussion below should be obvious to the reader. If we seek
the minimizer
of (\ref{20c}), we must satisfy
\be \bzed= {{\delta\Gamma_*}\over{\delta\bz(t)}}[\bz]= \bh[t;\bz]
                                + \sum_{k=1}^n
\bR_k^{-1}[\bz(t_k)-\br_k]\delta(t-t_k). \lb{20d} \ee
Thus, we see that $\bh[t;\bz_*]$ for the optimal $\bz_*[\br]$ must be a sum of
delta functions
at the observation times. This suggests that we consider only the estimation of
$\bz$ at the
observation times. In fact, we will see that this suffices.

The cost function
$H_*(\bz_1,...,\bz_n):=H_{Z,R}(\bz_1,...,\bz_n;\br_1,...\br_n)$ for estimating
$\bz_k:=\bz(t_k),
\,k=1,...,n$ is obtained in the following way. First, we define a {\it cumulant
generating function}
\be F_Z(\bolambda_1,...,\bolambda_n):= \log\langle\exp[\sum_{k=1}^n
\bolambda_k^\top \bZ(t_k)]\rangle. \lb{20e} \ee
This is entirely analogous to the cumulant generating functional $W_Z[\bh]$
defined in subsection 2.1.
In fact, they are equal with
\be \bh(t) = \sum_{k=1}^n \bolambda_k \delta(t-t_k). \lb{20f} \ee
The Legendre transform of $F_Z$ is the dynamical part of the cost function:
\be H_Z(\bz_1,...,\bz_n)= \max_{\bolambda_1,...,\bolambda_n}\left\{\sum_{k=1}^n
\bz_k^\top\bolambda_k
    -F_Z(\bolambda_1,...,\bolambda_n)\right\}. \lb{20g} \ee
This quantity is called the {\it multitime (relative) entropy}. It may also be
obtained via the Contraction
Principle directly from the effective action through a constrained
minimization:
\be  H_Z(\bz_1,...,\bz_n) = \min_{\{\bz:
\bz(t_k)=\bz_k,k=1,...,n\}}\Gamma_Z[\bz]. \lb{20h} \ee
The combined cost function is then
\be H_*(\bz_1,...,\bz_n)= H_Z(\bz_1,...,\bz_n)
    +   {{1}\over{2}}\sum_{k=1}^n [\br_k-\bz_k]^\top\bR_k^{-1}[\br_k-\bz_k].
\lb{20i} \ee
Its minimization yields the optimal values of $\bz(t_1),...,\bz(t_n)$. The
condition for the minimum is
\be \bolambda_k = \bR_k^{-1} [\br_k - \bz_k], \lb{20j} \ee
which can already be inferred from (\ref{20d}),(\ref{20f}). We may regard
(\ref{20j}) as a nonlinear
equation for either the $\bolambda$'s or the $\bz$'s.

To calculate numerically the cost function $H_Z(\bz_1,...,\bz_n)$ we see that
we must integrate the
forward and backward equations (\ref{6u}),(\ref{20r}) with a control field
$\bh(t)$ consisting of
delta-function spikes, as in (\ref{20f}). It is easiest to formulate this
integration in terms of
suitable {\it jump conditions} at the observation times. That is, we may
integrate the ordinary forward
and backward Kolmogorov equations (\ref{6n}),(\ref{6o}) with $\bh=\bzed$
between the observation times
but make discrete jumps at those times. We shall show that the proper jump
conditions are simply
\be \cP(\bx,t_k+) =
{{e^{\bolambda_k^\top\bcZ(\bx,t_k)}}\over{\cW(t_k-)}}\cP(\bx,t_k-), \lb{20k}
\ee
and
\be \cA(\bx,t_k-) =
{{e^{\bolambda_k^\top\bcZ(\bx,t_k)}}\over{\cW(t_k-)}}\cA(\bx,t_k+). \lb{20l}
\ee
Here we defined
\be \cW(t_k-):=\int d\bx\,\,e^{\bolambda_k^\top\bcZ(\bx,t_k)}\cP(\bx,t_k-)=
                                   \langle
e^{\bolambda_k^\top\bcZ(t_k)}\rangle_{t_k-}, \lb{20m} \ee
so that division by that factor guarantees proper normalization of the results
after the jump.

We prove now the validity of these jump conditions. For the first, it is useful
to make a reformulation
of the forward equation (\ref{6u}). The same solution found for that equation
may be obtained by solving instead
\be \partial_t \cQ(t) = \hat{L}(t)\cQ(t)+\bh^\top(t)\bcZ(t)\cdot \cQ(t)
\lb{20n} \ee
and then renormalizing subsequently
\be \cP(\bx,t):= {{\cQ(\bx,t)}\over{\cN(t)}} \lb{20o} \ee
with
\be \cN(t):=\int d\bx\,\,\cQ(\bx,t). \lb{20o2} \ee
It is not hard, by differentiating (\ref{20o}) with respect to time, to show
that $\cP(t)$ so-defined satisfies
(\ref{6u}). This is actually a standard device to solve the
Kushner-Stratonovich equation. In that context,
the analogue of equation (\ref{20n}) is called the {\it Zakai equation}
\cite{Zak}. With the delta-function
control field, we obtain
\be \partial_t \ln\cQ(t)= {{\hat{L}(t)\cQ(t)}\over{\cQ(t)}}
                 + \sum_{k=1}^n \bolambda_k \bcZ(t_k)\delta(t-t_k). \lb{20p}
\ee
We then integrate in time over the range $(t_k-\epsilon,t_k+\epsilon)$ and take
the limit as $\epsilon\rightarrow 0$.
The first term on the righthand side is continuous and does not contribute.
{}From the delta function contribution
we easily obtain
\be {{\cQ(\bx,t_k+)}\over{\cQ(\bx,t_k-)}} = e^{\bolambda_k^\top\bcZ(\bx,t_k)}.
\lb{20q} \ee
Renormalizing $\cQ(\bx,t_k+)$, we recover (\ref{20k}), as claimed.

% Likewise, the second jump condition can be best understood in terms of a
%%different variable than
% $\cC$ in (\ref{6v}). The proper variable is here, in fact, the original
%%adjoint variable $\cA=1+\cC$.
% It may be shown by a direct variational calculation, or by substituting
%%$\cA=1+\cC$ into (\ref{6v})
% that $\cA$ satisfies the equation
% \be \partial_t \cA(t)+
%%\hat{L}^*(t)\cA(t)+\bh^\top(t)[\bcZ(t)-\langle\bcZ(t)\rangle_t]\cA(t)=0.
%%\lb{20r1} \ee

The second jump condition can be similarly obtained. The backward equation
(\ref{20r}) must likewise be rewritten so that
the source terms stand alone before integration. In fact, by integrating
(\ref{20n}) over $\bx$, one finds that
\be {{d}\over{dt}}\ln\cN(t) = \bh^\top(t)\langle\bcZ(t)\rangle_t. \lb{20t} \ee
Using this result, one easily derives from (\ref{20r}) that
\be \partial_t \ln\left({{\cA(t)}\over{\cN(t)}}\right)
                       +
{{\hat{L}^*(t)\cA(t)}\over{\cA(t)}}+\bh^\top(t)\bcZ(t)=0. \lb{20s} \ee
Let us now consider the case where $\bh(t)$ is given by (\ref{20f}), as a sum
of delta-functions. Integrating (\ref{20s})
over the range $(t_k-\epsilon,t_k+\epsilon)$ and taking the limit as
$\epsilon\rightarrow 0$, then yields
\be {{\cA(\bx,t_k+)/\cN(t_k+)}\over{\cA(\bx,t_k-)/\cN(t_k-)}}=
e^{-\bolambda_k^\top\bcZ(\bx,t_k)}. \lb{20u} \ee
Of course, $\cN(t)$ itself experiences a jump across the observation time
$t_k$, changing as we have seen by the ratio
\be {{\cN(t_k+)}\over{\cN(t_k-)}}= \langle
e^{\bolambda_k^\top\bcZ(t_k)}\rangle_{t_k-}=\cW(t_k-). \lb{20v} \ee
This is a direct consequence of (\ref{20q}). From (\ref{20u}) and (\ref{20v}),
the second jump
condition immediately follows.

Using the jump conditions (\ref{20k}) and (\ref{20l}) to replace the controlled
forward and backward equations
(\ref{6u}),(\ref{20r}), the calculation of the cost function proceeds as
follows. Integrating (\ref{20t}) over
the time-interval $[t_i,t_f]$ and comparing with (\ref{6y}), we see that
\be F_Z(\bolambda_1,...,\bolambda_n)= \log \cN(t_f). \lb{20v1} \ee
By writing $\cN(t_f)= \prod_{k=1}^n {{\cN(t_k+)}\over{\cN(t_k-)}}$ (where
$\cN(t_1-)=1$ was used),
we can decompose this into a sum of contributions for each time $t_k$
\be F_Z(\bolambda_1,...,\bolambda_n) = \sum_{k=1}^n \,(\Delta
F)_k(\bolambda_1,...,\bolambda_k) \lb{20v2} \ee
with
\be (\Delta F)_k(\bolambda_1,...,\bolambda_k):= \log \langle
e^{\bolambda_k^\top\bcZ(t_k)}\rangle_{t_k-}=\log\cW(t_k-). \lb{20v3} \ee
Whereas the dependence upon $\bolambda_k$ is explicit, note that the dependence
upon the remaining variables
$\bolambda_1,...,\bolambda_{k-1}$ is only implicit through $\cP(\bx,t_k-)$.
Having determined $F_Z(\bolambda_1,...,\bolambda_n)$,
the entropy $H_Z(\bz_1,...,\bz_n)$ can then be obtained by the Legendre
transform formula (\ref{20g}).
In that formula $\bz_k := \bz(t_k),\,\,k=1,...,n$, with $\bz(t)$ given for all
times $t$ by
\be \bz(t) =  \int d\bx \,\,\bcZ(\bx,t)\cA(\bx,t)\cP(\bx,t). \lb{20z} \ee
It is worth emphasizing that the history $\bz(t)$ is a continuous function of
time. This will be true even though
the solutions $\cP(\bx,t),\cA(\bx,t)$ have jump discontinuities at the
observation times $t=t_k,\,\,k=1,...,n.$
In fact, it is easy to see by a direct differentiation that
\be {{d\bz}\over{dt}}(t) = <\{ \partial_t\hat{\bZ} +
[\hat{L}^*,\hat{\bZ}]\}\cA(t),\cP(t)>. \lb{20zz} \ee
In particular, all of the delta-function sources cancel from this equation.
Hence, $\bz(t)$ is continuous
but will generally have a time-derivative with jump-discontinuities.

The rest of the estimation protocols outlined in sections I.4 are the same. For
example, the double
minimization algorithm may be carried out using the Legendre dual pair
$H_Z(\bz_1,...,\bz_n),\,$
$F_Z(\bolambda_1,...,\bolambda_n)$. The iteration may be initiated by taking
\be \bz_k^{(0)}= \langle \bZ(t_k)\rangle_{t_k+} \lb{20w} \ee
and
\be \bolambda^{(0,0)}_k = \bR_k^{-1}[\br_k-\langle \bZ(t_k)\rangle_{t_k-}].
\lb{20x} \ee
The final result will be an optimal estimated history $\bz_*(t)$ given by
(\ref{20z}), where $\cP^*(t),\cA^*(t)$
therein are the solutions of the forward and backward equations for the
$\bolambda_k^*,\bz_k^*$ obtained as convergents
of the minimization algorithm.

It is worthwhile to compare this procedure for calculating the variational
estimator with discrete data to that for calculating the
optimal KSP estimator in the same circumstances. It is shown in Appendix 1 that
the optimal estimator may be obtained as well by
integrating the forward and backward Kolmogorov equations (\ref{6n}),(\ref{6o})
for $\cP_*(t),\cA_*(t)$ between the observation times
and by making discrete jumps at those times. The proper jump conditions are
\be \cP_*(\bx,t_k+) =
{{1}\over{\cW_*(t_k-)}}\exp\left[\bolambda_k^\top\bcZ(\bx,t_k)
-{{1}\over{2}}\delta\bcZ^\top(\bx,t_k-)
\bR_k^{-1}\delta\bcZ(\bx,t_k-)\right]\cP_*(\bx,t_k-), \lb{20z1} \ee
and
\be \cA_*(\bx,t_k-) =
{{1}\over{\cW_*(t_k-)}}\exp\left[\bolambda_k^\top\bcZ(\bx,t_k)
-{{1}\over{2}}\delta\bcZ^\top(\bx,t_k-)
\bR_k^{-1}\delta\bcZ(\bx,t_k-)\right]\cA_*(\bx,t_k+). \lb{20z2} \ee
In these equations
\be \bolambda_k = \bR_k^{-1}[\br_k-\langle \bZ(t_k)\rangle_{t_k-}], \lb{20z3}
\ee
the same as the zeroeth-order (\ref{20x}) above,
\be \delta\bcZ(\bx,t_k-): =\bcZ(\bx,t_k)-\langle \bZ(t_k)\rangle_{t_k-},
\lb{20z4} \ee
and
\be \cW_*(t_k-): =\int
d\bx\,\,\exp\left[\bolambda_k^\top\bcZ(\bx,t_k)
-{{1}\over{2}}\delta\bcZ^\top(\bx,t_k-)
\bR_k^{-1}\delta\bcZ(\bx,t_k-)\right]\cP(\bx,t_k-) \lb{20z5} \ee
is the factor to keep the density $\cP_*(\bx,t)$ normalized. The solutions of
these equations after a {\it single} forward and backward
integration then yield the conditional probability density, via the formula
$\cP(\bx,t|\cR(t_f))=\cA_*(\bx,t)\cP_*(\bx,t)$. See Appendix 1.

It is now clear that the zeroeth-order variational estimator, calculated after
one forward-backward sweep initialized with (\ref{20w}),(\ref{20x}),
does {\it not} coincide with the optimal KSP estimator, for the case of
discrete-time measurements. The main difference, one can easily see,
lies in the extra term in the exponent quadratic in $\delta\bcZ(\bx,t_k-)$.
This term has the effect of preventing the estimate after the jump from
being too far from the prior estimate $\langle \bZ(t_k)\rangle_{t_k-}$. The
absence of this term in the zeroeth-order variational equations helps
to make clear in what sense it is a ``mean-field'' approximation of the optimal
estimator, obtained by neglect of ``fluctuations''. In fact,
if we estimate not the variable $\bZ(t)$ itself, but rather its sample mean
$\overline{\bZ}_N(t)$, then heuristically $\bolambda$ remains
unchanged but $\delta\overline{\bZ}_N(t)= O(N^{-1/2})$. Hence, the quadratic
term in the exponent is $O(1/N)$ and may be
neglected. Of course, it must be realized that we are here comparing the
optimal KSP estimator with only a zeroeth-order approximation
to the variational estimator, not to the variational estimator itself for the
converged values of $\bolambda_k^*,\,\,k=1,...,n$. We hope that, in general,
the value of the variational estimator, calculated with the ``mean-field'' jump
rules (\ref{20k}),(\ref{20l}) for the minimizing values $\bolambda_k^*$,
will not be too far from the optimal KSP estimator given by the jump rules
(\ref{20z1}),(\ref{20z2}).

The difference between the zeroeth-order variational estimator and the optimal
KSP estimator which we have illustrated above for the case
of discrete-time measurements, of course also holds in the case of
continuous-time measurements. In that case it is a consequences of the
difference
in mathematical interpretation of the zeroeth-order variational equations and
the KSP stochastic equations, despite their formal identity.
Of course, at this point one might question the utility of the variational
formulation compared with the straightforward Bayesian approach
based upon the KSP equations. Whether in the discrete- or continuous-time
formulations, it is essentially just as difficult to solve the KSP
equations as to make one ``sweep'' in the iterative solution of the variational
problem. However, the latter requires in most cases a large number
of ``sweeps'' and furthermore provides a suboptimal estimate compared with the
KSP approach! The advantage of the variational approach will
become apparent in Part II, when we consider making finite-dimensional
approximations.

\newpage

\noindent {\bf I.6. Calculation of the Ensemble Dispersion}
% \subsection{Calculation of the Ensemble Dispersion}

\noindent As discussed in section I.1, one would like to have not just the mean
state $\bx_*(t)$ but also the covariance matrix $\bC_*(t)$
in the conditioned ensemble at each time $t$. As we shall show now, the
covariance may be readily calculated from the cost function itself.
For simplicity, we shall confine our discussion to the calculation of the
covariance of the measured variable $\bZ(t)$. The changes required for
the determination of the state covariance will be obvious.

Let us discuss first the case with discrete-time data. The entropy function
$H_*(\bz_1,...,\bz_n):=H_{Z,R}(\bz_1,...,\bz_n;
\br_1,...,\br_n)$ that we introduced in (\ref{20i}) of the last subsection has
also an interpretation as a generating function
for (irreducible) multitime correlations in the ensemble conditioned on
$\overline{\br}_N(t_j)=\br_j,\,\,j=1,...,n.$ Thus,
one may calculate the 2-time irreducible correlator
\be \boGamma_*(t_k,t_j) = \left.{{\partial^2
H_*}\over{\partial\bz_k\partial\bz_j}}(\bz_1,...,\bz_n)
                          \right|_{\bz=\overline{\bz}}. \lb{21a} \ee
This irreducible correlator is related to the multitime covariance matrix by
matrix inversion:
\be \bC_*(t_k,t_j) = [\boGamma_*(t_k,t_j)]^{-1}. \lb{21b} \ee
The same quantity could also be obtained from
$F_*(\bolambda^*_1,...,\bolambda^*_n):=F_{Z,R}
(\bolambda^*_1,...,\bolambda^*_n;\br_1,...,\br_n)$, the Legendre dual of
$H_*(\bz_1,...,\bz_n)$, as
\be \bC_*(t_k,t_j) = \left.{{\partial^2
F_*}\over{\partial\bolambda^*_k\partial\bolambda^*_j}}
(\bolambda^*_1,...,\bolambda^*_n)\right|_{\bolambda^*=\bzed}. \lb{21c} \ee
{}From this one may obtain the single-time covariance at any of the times $t_k$
by considering the diagonal
$\bC_*(t_k)=\bC_*(t_k,t_k)$. Without loss of generality, one may include any
time of interest as one of
the ``measurement times'' by simply taking the corresponding value of its
observation error as infinite,
or $\bR_k^{-1}=\bzed.$

While this procedure gives the correct result, it is not so practical because
the quantity of interest,
the diagonal $\bC_*(t_k)$, is obtained only through the intermediary of the
full 2-time covariance
$\bC_*(t_k,t_j)$. A more useful approach is based upon the single-time
generating function obtained from
the Contraction Principle
\be H_*(\bz;t_k):= \min_{\tilde{\bz}:\tilde{\bz}_k=\bz}
H_*(\tilde{\bz}_1,..,\tilde{\bz}_n). \lb{21d} \ee
This is just the {\it (conditional) relative entropy} at time $t_k$. We have
chosen here to make the time-dependence
explicit in the instantaneous entropy. One can calculate the Hessian of this
function
\be \boGamma_*(t_k) = \left.{{\partial^2
H_*}\over{\partial\bz\partial\bz}}(\bz;t_k)
                      \right|_{\bz=\overline{\bz}_k}. \lb{21e} \ee
and then obtain
\be \bC_*(t_k) = [\boGamma_*(t_k)]^{-1}. \lb{21f} \ee
To employ this method, one must carry out the minimization in (\ref{21d}). This
leads as before to the condition
\begin{eqnarray}
{{\partial H_*}\over{\partial\bz_j}}(\tilde{\bz}_1,...,\tilde{\bz}_n)
                                       & := &
\bolambda_j^*(\tilde{\bz}_1,...,\tilde{\bz}_n) \cr
                                    \, & =  &
\bolambda_j(\tilde{\bz}_1,...,\tilde{\bz}_n)
                                              +\bR_j^{-1}(\tilde{\bz}_j-\br_j)
\cr
                                    \, & =  & \bzed, \lb{21g}
\end{eqnarray}
for $j\neq k$ with $\tilde{\bz}_k=\bz$ fixed. This minimization problem can be
solved computationally
with the same methods used to find the global minimum, e.g. the double CG-type
algorithm, but now with $\tilde{\bz}_k=\bz$
held invariant and $H_*$ minimized only over the remaining variables
$\tilde{\bz}_j,\,j\neq k$. The result
will be the constrained minimizers $\bz^*_j(\bz;t_k)$ that, substituted into
$\bolambda_j^*(\tilde{\bz}_1,...,\tilde{\bz}_n)$
with $\tilde{\bz}_k=\bz$, give $\bzed$ for all $j\neq k$. However,
\be \bolambda_*(\bz;t_k) :=
\left.\bolambda_k^*(\tilde{\bz}_1,...,\tilde{\bz}_n)
\right|_{\tilde{\bz}_k=\bz;\,\,\tilde{\bz}_j=\bz^*_j(\bz;t_k),\,j\neq k}
\lb{21h} \ee
will not be zero. In fact, it is not hard to see that
\be \bolambda_*(\bz;t_k) = {{\partial H_*}\over{\partial\bz}}(\bz;t_k).
\lb{21i} \ee
Then, from (\ref{21e}),
\be \boGamma_*(t_k) =
\left.{{\partial\bolambda_*}\over{\partial\bz}}(\bz;t_k)
\right|_{\bz=\overline{\bz}_k}. \lb{21j} \ee
This gives rise to a simple, practical algorithm to calculate the covariance,
by means of a finite-difference
approximation for some small $\delta$
\begin{eqnarray}
 \Gamma_{\alpha\beta}^*(t_k) & \approx &
{{\lambda^*_\alpha(\bz^{+\beta};t_k)
-\lambda^*_\alpha(\bz^{-\beta};t_k)}\over{2\delta}} \cr
                          \, &    =    &
{{\lambda_\alpha(\bz^{+\beta};t_k)
-\lambda_\alpha(\bz^{-\beta};t_k)}\over{2\delta}}
            + [\bR^{-1}_k]_{\alpha\beta},
\lb{21k}
\end{eqnarray}
with
\be \bz^{\pm\beta}:= \overline{\bz}_k\pm \delta\cdot\hat{{\bf e}}_\beta,
\lb{21l} \ee
and $\hat{{\bf e}}_\beta$ a unit vector in the $\beta$-direction. We have set
$\bolambda(\bz;t_k) = \left.
\bolambda_k(\tilde{\bz}_1,...,\tilde{\bz}_n)\right|_{\tilde{\bz}_k=\bz;
\,\,\tilde{\bz}_j=\bz^*_j(\bz;t_k),\,j\neq k}$.
This approximation requires the calculation of $\bolambda_*(\bz;t_k)$ for the
two new values $\bz=\bz^{\pm\beta}$
displaced slightly from $\overline{\bz}_k$. This can be accomplished using
(\ref{21h}) and the double minimization
algorithm. Suitable guesses to initiate the minimization would be
\be \tilde{\bz}_j^{(0)}=\left\{ \begin{array}{ll}
                                 \bz^{\pm\beta}  & j=k \cr
                                 \overline{\bz}_j & j\neq k
                                \end{array} \right. \lb{21m} \ee
and
\be \bolambda^{(0,0)}_j = \bR_j^{-1}[\br_j-\tilde{\bz}_j^{(0)}],\,\,j=1,...,n.
\lb{21n} \ee
Of course, $\tilde{\bz}_k=\bz^{\pm\beta}$ is held fixed in the iteration. This
procedure must be followed to calculate
the covariance at each time $t_k$ of interest. For each scalar variable,
calculating its variance by this method
is roughly twice the work as calculating the optimal estimate itself over the
whole interval of time. However,
this statement is misleadingly pessimistic. In fact, the initial points
considered, $\tilde{\bz}_k=\bz^{\pm\beta},
\,\,\tilde{\bz}_j=\overline{\bz}_j,\,j\neq k$ are very close to the optimal
history, which is assumed known. Hence, only
small changes will occur in the $\tilde{\bz}_j$'s, $O(\delta)$ corrections to
the $\overline{\bz}_j$'s, and the minimization
algorithm should converge quite quickly. The contribution of the various small
changes can be read off from (\ref{21k}).
The direct contribution from the change in $\tilde{\bz}_k$ to $\bz^{\pm\beta}$
is $[\bC(t_k)]^{-1}+\bR_k^{-1}$,
where $\bC(t_k)$ is the covariance in the unconditioned ensemble. The
additional contributions from the small changes
in the $\tilde{\bz}_j,\,\,j\neq k$ will be similar, but will decay according to
the distance of $t_j$ from $t_k$ in time.
The rate of decay will be determined by some internal relaxation or memory time
of the system.

If the number of variables whose variance is required is large, then even the
matrix inversion in (\ref{21f}) is
difficult and should be avoided. This can be accomplished by following an
alternative procedure, based upon
implementing the constraint $\tilde{\bz}_k=\bz$ by a Lagrange multiplier. In
this case, (\ref{21d}) is replaced
by an unconstrained minimization
\be H_*(\bz;t_k):= \min_{\tilde{\bz}_1,...,\tilde{\bz}_n}
\tilde{H}(\tilde{\bz}_1,..,\tilde{\bz}_n;\tilde{\bolambda}), \lb{21o} \ee
where
\be \tilde{H}(\tilde{\bz}_1,..,\tilde{\bz}_n;\tilde{\bolambda}):
       =H_*(\tilde{\bz}_1,..,\tilde{\bz}_n) +
\tilde{\bolambda}^\top(\bz-\tilde{\bz}_k), \lb{21p} \ee
and the Lagrange multiplier $\tilde{\bolambda}$ is chosen subsequently to
impose the constraint $\tilde{\bz}_k=\bz$.
The condition for the minimum over all the variables
$\tilde{\bz}_j,\,\,j=1,...n,$ is
\begin{eqnarray}
{{\partial
\tilde{H}}\over{\partial\tilde{\bz}_j}}(\tilde{\bz}_1,...,\tilde{\bz}_n)
                             & :=  &
\bolambda_j^*(\tilde{\bz}_1,...,\tilde{\bz}_n)-\tilde{\bolambda}\delta_{jk} \cr
                          \, &  =  & \bzed. \lb{21q}
\end{eqnarray}
Thus, we see that the minimizing $\tilde{\bz}_j(\tilde{\bolambda};t_k)$'s in
(\ref{21o}) are nothing more than
\be \tilde{\bz}_j(\tilde{\bolambda};t_k)=
\left.\bz_j^*({\bolambda}_1^*,...,{\bolambda}_n^*)
\right|_{{\bolambda}_k^*=\tilde{\bolambda};\,\,{\bolambda}_j^*=\bzed,\,j\neq k}
\lb{21r} \ee
where
\be \bz_j^*({\bolambda}_1^*,...,{\bolambda}_n^*) :=
      {{\partial
F_*}\over{\partial\bolambda^*_j}}({\bolambda}_1^*,...,{\bolambda}_n^*).
\lb{21s} \ee
Then, using (\ref{21c}), one obtains
\begin{eqnarray}
\bC_*(t_k) & = &
\left.{{\partial\bz_k^*}\over{\partial\bolambda^*_k}}
({\bolambda}_1^*,...,{\bolambda}_n^*)
                 \right|_{{\bolambda}_j^*=\bzed,\,j=1,...,n} \cr
        \, & = &
\left.{{\partial\tilde{\bz}_k}\over{\partial\tilde{\bolambda}}}
(\tilde{\bolambda};t_k)
                 \right|_{\tilde{\bolambda}=\bzed}. \lb{21t}
\end{eqnarray}
Incidentally, it is clear also from (\ref{21r}) that the value of the Lagrange
multiplier to achieve the
constraint $\tilde{\bz}_k=\bz$ is just $\tilde{\bolambda}=\bolambda_*(\bz;t_k)$
as given in (\ref{21h}),(\ref{21i}).
The important point for our considerations here is that formula (\ref{21t})
involves no matrix inversion.

Thus, formula (\ref{21t}) becomes the basis of an alternative procedure to
numerically compute the covariance.
In this procedure, one minimizes
$\tilde{H}(\tilde{\bz}_1,..,\tilde{\bz}_n;\bolambda^{\pm\beta})$ for the values
\be \bolambda^{\pm\beta} = \pm \delta\cdot\hat{{\bf e}}_\beta \lb{21u} \ee
to obtain $\tilde{\bz}_j(\bolambda^{\pm\beta};t_k),\,\,j=1,...,n.$ Then one may
approximate
\be C^*_{\alpha\beta}(t_k) \approx
{{\tilde{z}_{k\alpha}(\bolambda^{+\beta};t_k)
-\tilde{z}_{k\alpha}(\bolambda^{-\beta};t_k)}\over{2
\delta}}. \lb{21v} \ee
The minimization to obtain the $\tilde{\bz}_j(\bolambda^{\pm\beta};t_k)$ may be
carried out with similar methods
as before, e.g. the double CG-type algorithm initiated with the guesses
\be \tilde{\bz}^{(0)}_j=\overline{\bz}_j,\,\,j=1,...,n \lb{21w} \ee
and
\be \bolambda^{(0,0)}_j=\bR_j^{-1}[\br_j-\overline{\bz}_j]
+\bolambda^{\pm\beta}\delta_{jk},\,\,j=1,...,n. \lb{21x} \ee
Our discussion above carries over straightforwardly to the case of
continuous-time data acquisition. The entropy
at any time $t_0$ is, by the Contraction Principle, given as
\be H_{Z,R}(\bz,\br;t_0) = \min_{\bz: \bz(t_0)=\bz} \Gamma_{Z,R}[\bz,\br]
\lb{21y} \ee
with $\Gamma_{Z,R}$ as in (\ref{11}). Alternatively, one has
\be H_{Z,R}(\bz,\br;t_0) = \min_{\bz}
\tilde{\Gamma}_{Z,R}[\bz,\br;\tilde{\bolambda}], \lb{21z} \ee
with
\be \tilde{\Gamma}_{Z,R}[\bz,\br;\tilde{\bolambda}]:=
\Gamma_{Z,R}[\bz,\br]+\tilde{\bolambda}^\top[\bz-\bz(t_0)].
\lb{21zz} \ee
Either of the approaches outlined above may be used to find $\bC(t_0;\br)$. For
example, in the second method
\be \bC(t_0;\br)=\left.{{\partial\tilde{\bz}}\over{\partial\tilde{\bolambda}}}
[t_0;\br,\tilde{\bolambda}]\right|_{\tilde{\bolambda}=\bzed} \lb{21yy} \ee
where $\tilde{\bz}[t;\br,\tilde{\bolambda}]$ is the solution of the
minimization condition
\be \bzed={{\delta\tilde{\Gamma}_{Z,R}}\over{\delta\bz(t)}}
[\bz,\br;\tilde{\bolambda}]
         = \bh[t;\bz,\br]-\tilde{\bolambda}\delta(t-t_0). \lb{21xx} \ee
This is solved with the double CG-type algorithm, using jump conditions
(\ref{20k}),(\ref{20l}) at $t_0$.

\section{Moment-Approximation of the Optimal Estimator}

\noindent {\bf II.1. The Rayleigh-Ritz Method}
% \subsection{The Rayleigh-Ritz Method}

\noindent Until now all of our theoretical work has been exact and without any
approximation, other than that involved in
conditioning on sample averages, the ``mean-field'' approximation discussed in
the section I.3. However, it is clear that
additional approximations are required to achieve a computationally tractable
estimation scheme for spatially-extended or
distributed systems. As discussed in the Introduction, the exact calculation of
the optimal nonlinear estimator by the KSP equations
is already known to be numerically unfeasible in such situations. Furthermore,
computation of the exact variational
estimator is just as impractical as the computation of the exact KSP optimal
estimator for a system with a large number of degrees-of-freedom.
It will not be possible for almost any system of real, practical interest. The
existence of a variational principle does not ameliorate
the basic computational difficulty imposed by the enormously many variables.
Just as for the KS filter, moment-closure appears to be
the only tractable numerical approach to an approximate solution. The advantage
of the variational formulation is that it permits
finite-dimensional approximations to be constructed by a Rayleigh-Ritz method
which preserves the main structural properties of the exact
estimator, discussed previously. We shall briefly discuss these features here,
referring to previous works \cite{Ey1,Ey3} for many details.

The Rayleigh-Ritz approximation to the cost function is obtained by means of
the characterization
of that functional through the constrained variation in (\ref{6q}). Rather than
varying over
all $\cA\in L^\infty,\cP\in L^1$, one varies only over finitely parametrized
trial functions.
The trial functions are constructed from the usual elements of a
moment-closure: a set of
{\it moment functions} $M_i(\bx,t),\,\,i=1,...,R$ and a PDF {\it Ansatz}
$\cP(\bx,t;\bomu)$,
which is conveniently parametrized by the mean values which it attributes to
the moment-functions,
$\bomu:=\int d\bx\,\,\cP(\bx,t;\bomu)\bM(\bx,t)$. The left trial function may
be taken to be
$\cA(t)= 1+[\cB(t)-\langle\cB(t)\rangle_t]$ with
\be \cB(\bx,t;\boalpha):= \sum_{i=1}^R \alpha_i M_i(\bx,t). \lb{21} \ee
Following the discussion in section 2.3, we have chosen the left trial state in
the form
(\ref{6t}), to incorporate automatically the overlap constraint (\ref{6r}). The
histories
$\boalpha(t),\bomu(t)$ are the parameters to be varied over. Substituting the
trial forms,
one obtains the reduced action
\be  \Gamma[\boalpha,\bomu]= \int_{t_i}^{t_f}
dt\,\,\boalpha^\top(t)[\dot{\bomu}(t)-\bV(\bomu(t),t)]
     \lb{22} \ee
with
\be \bV(\bomu,t):= \langle (\partial_t+\hL^*)\bM(t)\rangle_{\bomu(t)}. \lb{23}
\ee
Of course, $\langle\cdot\rangle_{\bomu(t)}$ denotes average with respect to the
PDF {\it Ansatz}.
An unconstrained variation of (\ref{22}) recovers the standard moment-closure
equation:
$\dot{\bomu}=\bV(\bomu,t)$. For the calculation of the action, however, there
is the additional
expectation constraint (\ref{6s}). In terms of the trial functions, it becomes
\be \bz(t)= \bozeta(\bomu(t),t)+ \bC_Z(\bomu(t),t)\boalpha(t). \lb{24} \ee
Here,
\be \bozeta(\bomu,t):= \langle\bZ(t)\rangle_\bomu \lb{25} \ee
is the $Z$-expectation within the PDF {\it Ansatz}
and
\be \bC_Z(\bomu,t):=
\langle\bZ(t)\bM^\top(t)\rangle_\bomu-\bozeta(\bomu,t)\bomu^\top \lb{26} \ee
is the corresponding $ZM$-covariance matrix. It is remarkable that
$\bozeta(\bomu,t),\bC_Z(\bomu,t)$
are the {\it only} inputs of the PDF {\it Ansatz} actually required for the
calculation.

When the constraint (\ref{24}) is incorporated into the action functional
(\ref{22}) by
means of a Lagrange multiplier $\bh(t)$, the resulting Euler-Lagrange equations
are
\begin{eqnarray}
\dot{\bomu} & = & \bV(\bomu,t)+ \bC_Z^\top(\bomu,t)\bh(t) \cr
         \, & := & \bV_Z(\bomu,\bh,t). \lb{27}
\end{eqnarray}
and
\be \dot{\boalpha}+\left({{\partial\bV_Z}\over{\partial\bomu}}\right)^\top
(\bomu,\bh,t)\boalpha
+\left({{\partial\bozeta}\over{\partial\bomu}}\right)^\top(\bomu,t)
\bh(t)=\bzed.
    \lb{28} \ee
These are solved subject to an initial condition $\bomu(t_i)=\bomu_0$ and a
final condition
$\boalpha(t_f)=\bzed$. When the solutions of the integrations are substituted
into (\ref{22}),
there results a Rayleigh-Ritz approximation $\tilde{\Gamma}_Z[\bz]$ to the
effective action
of $\bZ(t)$. The value $\bz(t)$ of the argument is that given by the constraint
equation (\ref{24})
for the given value of the control field $\bh(t)$. A corresponding
approximation of the
cumulant generating functional is given by
\be \tilde{W}_Z[\bh]= \int_{t_i}^{t_f}dt\,\,\bh^\top(t)\bozeta(\bomu(t),t)
\lb{29} \ee
in which $\bomu(t)$ is the solution of just the forward equation (\ref{27}) for
the
control history $\bh(t)$.

It is a very attractive feature of the above approximation scheme that the
resulting functionals
$\tilde{\Gamma}_Z[\bz],\tilde{W}_Z[\bh]$ remain formal Legendre transforms of
each other.
That is,
\be \tilde{W}_Z[\bh]+\tilde{\Gamma}_Z[\bz]= <\bh,\bz> \lb{30} \ee
and
\be \bh[t;\bz]={{\delta \tilde{\Gamma}_Z}\over{\delta\bz(t)}}[\bz],\,\,\,
    \bz[t;\bh]={{\delta \tilde{W}_Z}\over{\delta\bh(t)}}[\bh]. \lb{31} \ee
This fact makes it possible to carry over directly all of the minimax
algorithms
discussed in section I.4 for determination of optimal histories using the exact
cost
function to the Rayleigh-Ritz approximate one. Incidentally, the form of the
constraint (\ref{24}) makes it more apparent that this approach generalizes the
``sweep method'' employed in the case of linear dynamics \cite{Med}.

% \noindent {\it (4.2) Construction of the Approximate Estimator}

The iterative constructions that were discussed for the exact optimal estimator
can be followed also to calculate
the moment-closure approximation. For example, consider the two-step method. As
the first step,
one can calculate the approximate optimal $\tilde{\bz}_*[\br]$ given $\br$, by
minimizing
\be \tilde{\Gamma}_{Z,R}[\bz,\br]= \tilde{\Gamma}_{Z}[\bz]+
 {{1}\over{2}}\int_{t_i}^{t_f} dt\,\,[\br(t)-\bz(t)]^\top
\bR^{-1}(t)[\br(t)-\bz(t)].  \lb{16a} \ee
over $\bz$ with $\br$ fixed. This can be accomplished, for example, with a
double CG method
as before, taking now
\be \bz^{(0)}(t)= \bozeta(\bomu,t) \lb{18a} \ee
\be \bh^{(0,0)}(t)= \bR^{-1}(t)\left[\br(t)-\bozeta(\bomu,t)\right]  \lb{19a}
\ee
as the zeroth-order inputs. It is clear that (\ref{19a}), substituted into the
approximate
forward equation (\ref{27}), is formally equivalent to a moment-closure of the
KS-equation.
(Although it must be emphasized once more that, in the case of the KS filter,
the closure equation
analogous to (\ref{27}) must be regarded as a stochastic differential
equation.)
The second step is to calculate the approximate optimal $\tilde{\bx}_*[\bz]$ by
minimizing
$\tilde{\Gamma}_{X,Z}[\bx,\bz]$ over $\bx$ with $\bz$ fixed. Of course, the
Rayleigh-Ritz
approximation $\tilde{\Gamma}_{X,Z}[\bx,\bz]$ is calculated by the analogous
equations as
(\ref{27}),(\ref{28}):
\begin{eqnarray}
\dot{\bomu} & = & \bV(\bomu,t)+ \bC_X^\top(\bomu,t)\bk(t)
+\bC_Z^\top(\bomu,t)\bh(t) \cr
         \, & := & \bV_{X,Z}(\bomu,\bk,\bh,t). \lb{27a}
\end{eqnarray}
and
\be \dot{\boalpha}+\left({{\partial\bV_{X,Z}}\over{\partial\bomu}}\right)^\top
(\bomu,\bk,\bh,t)\boalpha
    +\left({{\partial\boxi}\over{\partial\bomu}}\right)^\top(\bomu,t)\bk(t)
+\left({{\partial\bozeta}\over{\partial\bomu}}\right)^\top
(\bomu,t)\bh(t)=\bzed, \lb{28a} \ee
where $\boxi(\bomu,t)$, $\bC_X(\bomu,t)$ are the closure $X$-mean and
$XM$-covariance, respectively.
The final approximate estimator is then the composition $\tilde{\bx}_*[\br]
=\tilde{\bx}_*[\tilde{\bz}_*[\br]]$.

However, there is a potential difficulty in applying the minimization
algorithms:
the Rayleigh-Ritz approximations to the cost functions need not be convex at
all!
Lack of convexity would correspond to a failure of realizability of the
predicted multi-time
correlations \cite{Ey1}. As a consequence of this failure, there might exist
local minima
in addition to the global one or, possibly, no minimum at all,
local or global. In the former case, a CG algorithm could be trapped in a local
minimum, and,
in the latter, it would not converge at all. Thus, for numerical purposes, it
is exceedingly
desirable to maintain convexity. It was shown in \cite{Ey2} that convexity will
be maintained
---at least for an expansion of the action to quadratic order in small
departures from the
minimum---whenever the relative entropy is a Lyapunov stability function for
the closure dynamics.
It is possible to construct closures for nonlinear stochastic dynamics which
guarantee the validity
of such an $H$-theorem \cite{GL}, using methods previously developed for
Boltzmann kinetic equations
in transport theory \cite{CDL}. One example of the general scheme are closures
based upon an exponential
PDF {\it Ansatz}. Such closures have the property that the relative entropy
satisfies an $H$-theorem
and thus (local) convexity of the Rayleigh-Ritz approximations is guaranteed.
This is discussed
further in \cite{GL} and in section II.3 below.

\newpage

% \noindent {\it (4.3) Discrete-Time Data and Ensemble Dispersion}

\noindent {\bf  II.2. Discrete-Time Data and Ensemble Dispersion}
% \subsection{Discrete-Time Data and Ensemble Dispersion}

\noindent We have seen in section I.5. that the estimation problem based upon
discrete-time data has,
in the exact formulation, a simple solution in terms of certain jump
conditions. The situation is worse
for closure approximations. In fact, we shall see below that, for general
moment-closures, the approximate
``smoother'' with discrete-time data may not even be continuous at the
observation times!
This is an important failing since these same methods are also involved in the
calculation
of instantaneous ensemble dispersions, as we have seen in section I.6.

Let us illustrate the nature of the problem for a general moment-closure. If
one differentiates
the expression for $\bz(t)$ in (\ref{24}) with respect to time, using the
variational equations (\ref{27}),
(\ref{28}), simple computations give a result of the form
\begin{eqnarray}
    {{dz_a}\over{dt}}(t) & = &  {{d\zeta_a}\over{dt}}(\bomu,t) + \alpha_j
    \left[{{dC^Z_{aj}}\over{dt}}(\bomu,t)-C^Z_{ai}(\bomu,t){{\partial
V_j}\over{\partial\mu_i}}(\bomu,t)\right] \cr
                     \, &   &  \,\,\,\,\,\,\,+
h_b(t)\left[C^Z_{bi}(\bomu,t){{\partial\zeta_a}\over{\partial\mu_i}}(\bomu,t)-
C^Z_{ai}(\bomu,t){{\partial\zeta_b}\over{\partial\mu_i}}(\bomu,t)\right] \cr
                     \, &   &  \,\,\,\,\,\,\,+
    h_b(t)\left[C^Z_{bi}(\bomu,t){{\partial
C^Z_{aj}}\over{\partial\mu_i}}(\bomu,t)-
                C^Z_{ai}(\bomu,t){{\partial
C^Z_{bj}}\over{\partial\mu_i}}(\bomu,t)\right]\alpha_j. \lb{33}
\end{eqnarray}
For any function of $\bomu,t$ we set ${{d}\over{dt}}:=
{{\partial}\over{\partial t}} + \bV(\bomu,t)\bdot\grad_\bomu.$
Equation (\ref{33}) should be compared with the exact result in (\ref{20zz}).
In contrast to the cancellation of the
explicit $\bh(t)$ terms found there, such terms remain in the second and third
lines above. Only in the case where
a single scalar variable $z(t)$ is considered and thus $a=b=1$ is there an
obvious cancellation in the last two terms.
This means that, in general, the delta-functions will not cancel if one
considers a control field of the form
$\bh(t)= \sum_k \bolambda_k\delta(t-t_k)$, as appropriate for discrete-time
data, and $\bz(t)$ itself
will have jump-discontinuities at the measurement times $t_k$. Of course, one
take observations, not instantaneously,
but instead averaged over a small interval of time $\tau$. The delta functions
are then replaced by approximate
delta's $\delta_\tau(t-t_k)$ with time-window $\tau$. However, the problem will
reappear when $\tau$ is taken very small,
for then $\bz(t)$ will change sharply at times $t_k$.

A related problem has to do with the formulation of proper jump conditions in
the same circumstances. Let us
even assume that $z(t)$ is a single scalar. Then, the forward equation for the
moment variable $\bomu$ becomes
\be  \dot{\mu}_i = V_i(\bomu,t) + h(t) C^Z_i(\bomu,t). \lb{34} \ee
If $h(t)$ is a sum of delta-functions, then one cannot integrate the equation
to obtain the jumps in $\mu_i$
at the observation times. The difficulty is that $C^Z_i(\bomu,t)$ will then
also have jump-discontinuities at
those times and it is impermissable to integrate a delta-function against a
discontinuous function. The obvious
strategy is first to divide both sides by $C^Z_i(\bomu,t)$ and only afterward
integrate across the jump. There is
still a problem however. The other variables besides $\mu_i$ in the integrand
also make jumps and it is therefore
ambiguous which value should appear as the integration range shrinks to zero.
Thus, the strategy only works in the
case where there is also a single scalar moment variable $\mu(t)$. In that
case, we can integrate and obtain a jump
condition in the form of an ``area rule'':
\be \int_{\mu_k^-}^{\mu_k^+} {{d\mu}\over{C^Z(\mu,t_k)}} = \lambda_k. \lb{35}
\ee
We have set $\mu_k^\pm = \mu(t_k\pm)$. The backward jump condition for the
adjoint variable $\alpha$
then follows most easily from the continuity of $z(t)$ noted above. With
notations as above, $\alpha_k^\pm
= \alpha(t_k\pm)$ and so forth, we have
\be \zeta^+_k + C^{Z+}_k\alpha^+_k = \zeta^-_k + C^{Z-}_k\alpha^-_k. \lb{36}
\ee
The jumps in $\zeta,C^Z$ are known, because these are assumed continuous
functions of $\mu,t$ and the
jump in $\mu$ is known from (\ref{35}). Solving for the backward jump gives
\be \alpha^-_k = {{(\zeta_k^+-\zeta_k^-) + C^{Z+}_k
\alpha^+_k}\over{C^{Z-}_k}}. \lb{37} \ee
Hence, only in the case of a single scalar moment function and observation
variable is it obvious how
to formulate jump conditions, in the case of a general moment-closure.

It still remains in that case to formulate the algorithm to calculate the cost
function itself.
The proper definition turns out to be
\be F_Z(\lambda_1,...,\lambda_n):= \sum_{k=1}^n (\Delta
F)_k(\lambda_1,...,\lambda_k) \lb{37a} \ee
where the increment at time $t_k$ is given by a second area rule:
\be \int_{\mu_k^-}^{\mu_k^+} {{\zeta(\mu,t_k)}\over{C^Z(\mu,t_k)}}\,\,d\mu =
(\Delta F)_k. \lb{37b} \ee
There is a simple heuristic motivation for both this rule and the previous one.
In fact, the basic approximation
is to replace the exponentially-modified PDF by a PDF from the {\it Ansatz}
with an adjusted moment. That is,
\be {{1}\over{\cW(\lambda_k;\mu_k^-,t_k)}}e^{\lambda_k
Z(\bx,t_k)}P(\bx,t_k;\mu_k^-) \approx P(\bx,t;\mu_k^+),
    \lb{37c} \ee
with the normalization factor
\be \cW(\lambda_k;\mu_k^-,t_k):= \int d\bx\,\,e^{\lambda_k
Z(\bx,t_k)}P(\bx,t_k;\mu_k^-). \lb{37d} \ee
The $M$-moment of (\ref{37c}) is
\begin{eqnarray}
\mu_k^+(\lambda_k;\mu_k^-,t_k) & := &
{{1}\over{\cW(\lambda_k;\mu_k^-,t_k)}}\int d\bx\,\,M(\bx,t_k) e^{\lambda_k
Z(\bx,t_k)}P(\bx,t_k;\mu_k^-) \cr
                           \,  & \approx & \mu_k^+. \lb{37e}
\end{eqnarray}
Differentiating once and using again (\ref{37c}) thus gives
\be {{\partial \mu_k^+}\over{\partial \lambda_k}}(\lambda_k;\mu_k^-,t_k) =
C^{Z}(\mu_k^+,t_k). \lb{37f} \ee
The first area rule (\ref{35}) is just an integral form of this latter relation
(\ref{37f}). Likewise,
if we define
\be (\Delta F)_k(\lambda_k;\mu_k^-,t_k):= \log \cW(\lambda_k;\mu_k^-,t_k),
\lb{37g} \ee
then we see by applying (\ref{37c}) twice again that
\be {{\partial (\Delta F)_k}\over{\partial \lambda_k}}(\lambda_k;\mu_k^-,t_k)=
\zeta(\mu_k^+,t_k). \lb{37h} \ee
The second area rule (\ref{37b}) is likewise the integral form of (\ref{37h}).
Note from (\ref{37g}) that all
of the dependence of $(\Delta F)_k$ upon $\lambda_1,...,\lambda_{k-1}$ is
through $\mu_k^-$, analogous to (\ref{20v3}).
The cost function $H_Z(z_1,...,z_n)$ is finally defined as the Legendre dual of
 $F_Z(\lambda_1,...,\lambda_n)$ given
by (\ref{37a}). The jump conditions (\ref{35}),(\ref{37}) may be used very much
as the exact ones (\ref{20k}),(\ref{20l})
for the purposes of estimation with discrete-time data and of ensemble variance
calculation. Only the
Rayleigh-Ritz approximations of the cost functions need be substituted for the
exact ones in the algorithms
described earlier. In calculating the Legendre dual $H_Z(z_1,...,z_n)$ the
adjoint equation is used to evaluate
$z_k = \zeta_k^\pm + C_k^{Z\pm}\alpha_k^\pm$. Of course, one should check that
this gives the same result
as direct differentiation $z_k= {{\partial F_Z}\over{\partial\lambda_k}}$. This
is true but we shall not give the proof here,
because we prove a very similar result in the Appendix 2. The proof is based
upon the easily established relations
\be {{\partial\mu_k^+}\over{\partial\mu_k^-}}= {{C^{Z+}}\over{C^{Z-}}},\,\,\,\,
    {{\partial (\Delta F)_k}\over{\partial\mu_k^-}}=
{{\zeta_k^+-\zeta_k^-}\over{C^{Z-}}}. \lb{37i} \ee
Of course, the adjoint equation need not be employed at all, but it is a
convenient way of evaluating
the required derivative.

Substantial simplifications in the jump conditions occur in the important
special case where $Z=M$.
In that case $\zeta(\mu,t)=\mu,\,\,C^Z(\mu,t)=C(\mu,t)$. We can then define a
function
\be \lambda(\mu,t):= \int_{\mu(t)}^\mu \,\,{{d\bar{\mu}}\over{C(\bar{\mu},t)}}
\lb{37j} \ee
where $\mu(t)$ is the solution of the {\it unperturbed} moment equation. If
$\mu(\lambda,t)$
is the inverse function, then we can also define
\be F(\lambda,t):= \int_0^\lambda \mu(\bar{\lambda},t)\,\,d\bar{\lambda}.
\lb{37k} \ee
It follows by our definitions that
\be F'(\lambda,t)=\mu,\,\,F''(\lambda,t)=C. \lb{37l} \ee
In terms of the function $\lambda(\mu,t)$ the first area rule (\ref{35})
becomes
\be \lambda(\mu_k^+,t_k)-\lambda(\mu_k^-,t_k)=\lambda_k. \lb{37m} \ee
Also, using (\ref{37l}) we note that
\begin{eqnarray}
(\Delta F)_k  & = & \int_{\mu_k^-}^{\mu_k^+}\,\,
                    {{\mu \,\,d\mu}\over{\left({{d\mu}\over{d\lambda}}\right)}}
\cr
           \, & = & \int_{\lambda_k^-}^{\lambda_k^+}\mu(\lambda)\,\,d\lambda
\cr
           \, & = & F(\lambda_k^+,t_k)-F(\lambda_k^-,t_k). \lb{37n}
\end{eqnarray}
Hence, the ``area rules'' are replaced by equations involving discontinuities
of explicit functions,
always assuming, of course, that integrals defining the functions in
(\ref{37j}),(\ref{37k}) may be evaluated.
The key to this simplification was the relations in (\ref{37l}), which imply
that $F$ is a convex ``potential''
generating the first and second moments of the PDF {\it Ansatz}. Such a
potential will always exist
for functions of one variable, but not in general for multivariate functions.

\noindent {\bf II.3. Exponential PDF Closures}
% \subsection{Exponential PDF Closures}

\noindent We have seen above that, for a general closure, there is a
satisfactory treatment of estimation with discrete-time data {\it only}
for the case where there is both a single measured variable $Z(t)$ and a single
closure variable $M(t)$. Obviously, this is an extreme
limitation. However, it may be possible to circumvent this severe restriction
within special classes of closures. In fact, as we show now,
closures constructed with an exponential PDF {\it Ansatz} have better
properties.  We shall see that they guarantee continuity of optimal estimators.
Furthermore, they provide very simple ``jump-conditions'' for estimation with
discrete-time data.

Exponential PDF closures are one example of the general class considered in
\cite{GL}. Hence, we shall only
make a quick summary of the properties required here and refer the interested
reader to the paper \cite{GL}
for more details. Most concretely, the class of closures we consider are those
built from a PDF {\it Ansatz}
of the exponential form:
\be \cP(\bx,t;\bolambda) =
{{\exp(\bolambda^\top\bM(\bx,t))}\over{\cN(\bolambda,t)}}\cP_*(\bx,t) \lb{38}
\ee
with
\be \cN(\bolambda,t):= \int d\bx\,\,\exp(\bolambda^\top\bM(\bx,t))\cP_*(\bx,t).
\lb{39} \ee
Here $\cP_*(\bx,t)$ is a {\it reference PDF}. To guarantee some of the good
properties of the closure
discussed in \cite{GL}, the reference PDF must be a solution (or approximate
solution) of the Fokker-Planck
equation. However, for the properties discussed here, $\cP_*(t)$ may be an
arbitrary PDF. The exponential
family in (\ref{38}) is parameterized by the ``potential'' variables
$\bolambda$, rather than by the moments
$\bomu$ of the closure variables $\bM(\bx,t)$. However, there are simple
relationships between these quantities.
We may define
\be  F(\bolambda,t):= \log \cN(\bolambda,t), \lb{40} \ee
which is a cumulant-generating function for the variables $\bM(t)$ in the PDF
{\it Ansatz}. Likewise, its Legendre transform
\be  H(\bomu,t): = \max_\bolambda \left\{\bomu^\top\bolambda -
F(\bolambda,t)\right\} \lb{41} \ee
is a generating function for irreducible correlation functions of $\bM(t)$. It
is the {\it relative entropy}
for the PDF {\it Ansatz} in (\ref{38}) with respect to the reference PDF
$P_*(t)$. Under some conditions
discussed in \cite{GL}, it satisfies an H-theorem for the closure dynamics
constructed with the {\it Ansatz}.
However, the role of $F,H$ as generating functions will be more important for
us here. Thus, $\bomu={{\partial F}
\over{\partial\bolambda}}$ and conversely $\bolambda={{\partial
H}\over{\partial\bomu}}$. It is a consequence
of the former that ${{\partial\bomu}\over{\partial\bolambda}}$ is the
covariance matrix $\bC$ of $\bM$ and that
${{\partial\bC}\over{\partial\bolambda}}$ is the 3rd-order cumulant. These
relationships will prove to be important
in the following.

We shall now show that, for the exponential PDF closures, the history $\bz(t)$
is continuous even for $\bh(t)$
consisting of delta-function spikes, {\it when the variables} $\bcZ(t)$ {\it
are among the closure variables}
$\bM(t)$ {\it themselves}. This last condition places some restriction, but a
fairly modest and natural one.
Without any loss of generality, we can consider $\bcZ(t)$ to consist of the
entire set of closure variables $\bM(t)$.
As before, some of our previous formulas then simplify considerably. For
example, $\bozeta(\bomu,t)
=\bomu$ and $\bC^Z(\bomu,t)=\bC(\bomu,t)$, the usual $MM$-covariance matrix.
Then (\ref{24}) is replaced by
\be \bm(t)= \bomu(t) + \bC(\bomu(t),t)\boalpha(t). \lb{42} \ee
The time-derivative of the latter, given in general in (\ref{33}), also
simplifies. In fact, the term in the
bracket in the second line becomes
\be C_{ba}(\bomu,t)-C_{ab}(\bomu,t) = 0 \lb{43} \ee
which vanishes by the symmetry of the covariance matrix. The term in the
bracket in the third line of (\ref{33})
becomes
\be C_{ajb}(\bomu,t)- C_{bja}(\bomu,t)= 0 \lb{44} \ee
where $C_{abc}(\bomu,t)$ is the 3rd-order cumulant of $\bM(t)$. Indeed,
\be    {{\partial C_{aj}}\over{\partial\mu_i}}
       = {{\partial
C_{aj}}\over{\partial\lambda_k}}{{\partial\lambda_k}\over{\partial\mu_i}}
       = C_{ajk}\Gamma_{ki}. \lb{45} \ee
Since the irreducible 2nd correlator is the inverse covariance matrix,
$\boGamma=\bC^{-1}$, the expression
in (\ref{44}) follows from the corresponding expression in (\ref{33}). However,
it is obvious that (\ref{44})
vanishes, by the symmetry of the 3rd-order cumulant. Putting together all of
these results, we have
\be {{d\bm}\over{dt}}(t) = \bV(\bomu,t)+
                  \left[
{{d}\over{dt}}\left({{\partial\bomu}\over{\partial\bolambda}}\right)
                         -
\left({{\partial\bV}\over{\partial\bolambda}}\right)^\top \right]\boalpha.
\lb{46} \ee
This should be compared with the exact expression (\ref{20zz}). Just as there,
we see that the terms directly
involving $\bh(t)$ all cancel. Hence, $\bm(t)$ remains continuous even with
$\bh(t)$ containing delta-function spikes.

We shall finally show that the exponential PDF closures also permit the
formulation of simple jump conditions
at the times $t_k$ where the delta functions occur. This should not be too
surprising, when one considers that
the exact jump conditions in (\ref{20k}),(\ref{20l}) consist simply of suitable
exponential modifications of the
solutions of the foward, backward equations.  To derive the jump conditions in
the closure, we use a strategy
motivated by that in section II.2. In fact, observe by
$\bC={{\partial\bomu}\over{\partial\bolambda}}$
and the chain rule that $\dot{\bomu}=\bC(\bolambda)\dot{\bolambda}.$ Thus, if
one defines $\bW(\bolambda):=
\boGamma(\bomu)\bV(\bomu)$ and $\bogamma:=\bC(\bolambda)\boalpha,$ then in
terms of the new variables $\bogamma,\bolambda$,
the nonequilibrium action, including the constraint term with the Lagrange
multiplier, becomes
\be \Gamma[\bogamma,\bolambda]= \int_{t_i}^{t_f} dt\,\,\left\{
\bogamma^\top[\dot{\bolambda}-\bW(\bolambda)]
     +\bh^\top(t)[\bm(t)-\bomu(\bolambda)-\bogamma]\right\}. \lb{46a} \ee
The Euler-Lagrange equations in terms of these variables become
\be \dot{\bolambda} = \bW(\bolambda,t) + \bh(t), \lb{48} \ee
\be \dot{\bogamma}+
\left({{\partial\bW}\over{\partial\bolambda}}\right)^\top(\bolambda,t)\bogamma
                                                                    +
\bC(\bolambda,t)\bh(t)=\bzed, \lb{48a} \ee
and the constraint equation
\be \bm(t) = \bomu(\bolambda,t)+ \bogamma. \lb{48b} \ee
In the first equation (\ref{48}) we may integrate across the spike
$\bolambda_k\delta(t-t_k)$ in $\bh(t)$ to obtain
\be \bolambda(\bomu_k^+,t_k)-\bolambda(\bomu_k^-,t_k)= \bolambda_k. \lb{49} \ee
These are the {\it forward jump conditions}. As should not be unexpected, the
potential $\bolambda(\bomu,t)$
is simply incremented by $\bolambda_k$ at the spike. A similar result can be
obtained by integrating the
backward closure equation (\ref{48a}) across the spike. However, it is simpler
to use the continuity of $\bm(t)$
at the jump, which was established above.  Then from (\ref{48b}) one
immediately derives
\be \bogamma^-_k = (\bomu_k^+-\bomu_k^-) + \bogamma^+_k. \lb{51} \ee
These are the {\it backward jump conditions}.

The multi-time cumulant-generating function $F(\bolambda_1,...,\bolambda_n)$
can be obtained from (\ref{20v1})
with the observation that $\cN(t_f)= \prod_{k=1}^n
{{\cN(\bolambda_k^+,t_k)}\over{\cN(\bolambda_k^-,t_k)}}$
and thus (\ref{20v2}) holds with
\begin{eqnarray}
(\Delta F)_k(\bolambda_1,...,\bolambda_k)  & = &
F(\bolambda_k^+,t_k)-F(\bolambda_k^-,t_k) \cr
           \, & = & F(\bolambda_k^-+\bolambda_k,t_k)-F(\bolambda_k^-,t_k),
\lb{52}
\end{eqnarray}
generalizing (\ref{37n}). Then the multi-time entropy $H(\bm_1,...,\bm_n)$ is
obtained by the Legendre transform
\be H(\bm_1,...,\bm_n)= \sum_{k=1}^n \bm_k^\top\bolambda_k -
F(\bolambda_1,...,\bolambda_k). \lb{53} \ee
with $\bm_k$ given by (\ref{42}), $\bm_k=\bm(t_k)$, for $t=t_k,\,\,k=1,...,n$.
Of course, it must be shown that
\be \bm_k={{\partial F}\over{\partial\bolambda_k}} \lb{53a} \ee
for all $k=1,...,n$ in order for (\ref{53}) to be valid. Cf. equation
(\ref{20g}). The proof is somewhat technical,
so it is given in the Appendix 2.

While the previous approximation has a rather elegant and tractable
formulation, there is nevertheless also an
unpleasant asymmetry between forward and backward time directions. Thus,
information propagates forward in time
via the nonlinear closure equation (\ref{48}), but information propagates
backward in time via the
equation (\ref{48a}) which is linear in the adjoint variable $\bogamma$.
Ultimately, this asymmetry is due to our
employment of a nonlinear (exponential) {\it Ansatz} (\ref{38}) for the PDF,
while the solution of the backward equation
is taken to be of the linear form (\ref{21}). However, there is nothing in the
Rayleigh-Ritz method which requires
the use of the linear {\it Ansatz} (\ref{21}) for the left trial state. In
fact, that expression has other unpleasant
features. The exact backward Euler-Lagrange equation (\ref{20r}) is known to be
positivity-preserving, so that
the solution $\cA(\bx,t)$ starting from final data $\cA(t)\equiv 1$ must be
everywhere nonnegative. However, the
linear {\it Ansatz} $\cA(\bx,t)= 1 + \sum_{i=1}^R \alpha_i(t)
[M_i(\bx,t)-\mu_i(t)]$ may easily become negative,
if the adjoint variables $\boalpha$ become large enough in magnitude. It is
therefore desirable to consider more
general {\it Ans\"{a}tze} for the left trial state than the linear one.

Within the context of exponential PDF closures a particularly symmetric and
attractive choice is to make
the {\it double exponential Ansatz}:
\be \cP(\bx,t) =
\exp\left[\bobeta^\top\bM(\bx,t)-F(\bobeta,t)\right]\cP_*(\bx,t) \lb{53b} \ee
for the right trial state and
\be \cA(\bx,t) = \exp\left[\boalpha^\top\bM(\bx,t)-(\Delta_\boalpha
F)(\bobeta,t) \right] \lb{53c} \ee
for the left trial state. Here $(\Delta_\boalpha F)(\bobeta,t):=
F(\boalpha+\bobeta,t)-F(\bobeta,t)$ so that the
normalization constraint $<\cA(t),\cP(t)>=1$ is automatically satisfied. It is
then clear that, for small $\boalpha$,
(\ref{53c}) coincides with the linear {\it Ansatz}. (Note that
$(\Delta_\boalpha F)(\bobeta,t)=\boalpha^\top\bomu(\bobeta,t)
+O(\alpha^2)$.) However, this new {\it Ansatz} is globally nonnegative and
symmetric in form to the exponential for
the right trial state. An even more attractive feature of this double
exponential {\it Ansatz} is that, within it,
the Rayleigh-Ritz effective action of the closure variables $\bM$ themselves
may be calculated analytically in closed form.
The result is:
\be \Gamma[\bm] = {{1}\over{4}} \int_{t_i}^{t_f}
dt\,\,[\dot{\bm}(t)-\bV(\bm,t)]^\top
                                \bQ^{-1}(\bm,t) [\dot{\bm}(t)-\bV(\bm,t)],
\lb{53d} \ee
where
\be Q_{ij}(\bm,t) := \langle (\grad_\bx M_i)^\top \bD (\grad_\bx
M_j)\rangle_{\bolambda(\bm,t)}. \lb{53e} \ee
This effective action has precisely the Onsager-Machlup form. The statement
generalizes a previous result in \cite{Ey2},
for general closures, that the Rayleigh-Ritz effective action has the
Onsager-Machlup form to quadratic order.
Let us just briefly sketch the derivation, which will be given in detail
elsewhere \cite{Ey4}, along with a complete discussion
of its remarkable properties. It is a straightforward calculation to show that
\be (\partial_t + \hL^*)\cA(\bx,t) = \left\{
\dot{\boalpha}^\top\bM(\bx,t)+\boalpha^\top\dot{\bM}(\bx,t)
    +\grad_\bx(\boalpha^\top\bM)\bdot\bD\bdot\grad_\bx(\boalpha^\top\bM)
-\Delta_\boalpha\dot{F}(\bobeta,t)\right\}\cA(\bx,t). \lb{53f} \ee
In that case
\be <(\partial_t + \hL^*)\cA(t),\cP(t)> =
\dot{\boalpha}^\top\bomu(\bolambda,t)+\boalpha^\top \bV(\bolambda,t)
    +  \boalpha^\top \bQ(\bolambda,t)\boalpha
-\Delta_\boalpha\dot{F}(\bobeta,t), \lb{53g} \ee
where
% \be \bolambda:=\boalpha+\bobeta. \lb{53h} \ee
$ \bolambda:=\boalpha+\bobeta.$
However, it is easy to see that the second constraint on the mean values
becomes in these variables
% \be \bm(t)=<\cA(t),\bM(t)\cP(t)> = \bomu(\bolambda,t). \lb{53i} \ee
$\bm(t)=<\cA(t),\bM(t)\cP(t)> = \bomu(\bolambda,t).$
Thus, holding the history $\bm(t)$ fixed is equivalent to holding
$\bolambda(t)$ fixed. We cannot vary independently
% over $\boalpha(t)$ and $\bobeta(t)$, but one is determined from the other via
%%the relation (\ref{53h}). Since, for fixed $\bm(t)$,
over $\boalpha(t)$ and $\bobeta(t)$, but one is determined from the other via
the relation $\bolambda(t)=\boalpha(t)+\bobeta(t)$.
Since, for fixed $\bm(t)$, (\ref{53g}) implies that
\be \Gamma[\boalpha,\bobeta] =
\int_{t_i}^{t_f}dt\,\,\left\{\boalpha^\top[\dot{\bm}-\bV(\bm,t)]
                               -\boalpha^\top\bQ(\bm,t)\boalpha\right\},
\lb{53j} \ee
maximizing over $\boalpha$ yields (\ref{53d}).

Although the Onsager-Machlup form (\ref{53d}) is most interesting for theory,
practical estimation is easier with the
latter expression (\ref{53j}). Including the cost function for the
observations, the total action to be minimized is
\be \Gamma_*[\boalpha,\bm] =
\int_{t_i}^{t_f}dt\,\,\left\{\boalpha^\top[\dot{\bm}-\bV(\bm,t)]
                               -\boalpha^\top\bQ(\bm,t)\boalpha\right\}
                             +{{1}\over{2}}\sum_{k=1}^n
[\bm(t_k)-\br_k]^\top\bR_k^{-1}[\bm(t_k)-\br_k].
\lb{53k} \ee
The Euler-Lagrange equations of this problem are
\be \dot{\bm} = \bV(\bm,t) + 2\bQ(\bm,t) \boalpha, \lb{53l} \ee
\be \dot{\boalpha} +
\left({{\partial\bV}\over{\partial\bm}}\right)^\top\boalpha
    + {{\partial}\over{\partial\bm}}\left(\boalpha^\top\bQ\boalpha\right) =
    \sum_{k=1}^n \bR_k^{-1}[\bm(t_k)-\br_k]\delta(t-t_k). \lb{53m} \ee
Solving these equations with boundary values $\bm(t_i)=\bm_0$ and
$\boalpha(t_f)=\bzed$ can give directly
the optimal history, without the need of applying any explicit minimization
algorithm. It is transparent
in this formulation that the optimal history $\bm_*(t)$ is continuous at the
observation times, because
the first equation (\ref{53}) contains no delta-functions in time. Only the
adjoint variables $\boalpha(t)$
suffer jumps at the measurement times $t=t_k$.

The same circle of ideas may be applied to constructing closures of the KSP
equations
for the optimal history itself, rather than just the variational approximation.
In fact, assume that
the closure variables $\bM$ consist of the measured variables $\bZ$ and their
tensor products $\bZ\otimes\bZ$,
$\bM:=(\bZ,\bZ\otimes\bZ)$, with mean values given by $\bm=(\bozeta,\boZeta)$
for a double exponential Ansatz.
Let the exponential parameters in the left trial state then be denoted as
$(\boalpha,{\bf A})$ and those in the
right trial state as $(\bobeta,{\bf B})$. Because the states evolve by the
(unperturbed) forward and
backward Kolmogorov equations between measurements, the Euler-Lagrange
equations within the closure
are of the same form as those in (\ref{53l}),(\ref{53m}). As there, there are
no jumps at measurement
times in the equations for $\bm=(\bozeta,\boZeta).$ On the other hand, there
are simple jump conditions
for the adjoint variables $(\boalpha,{\bf A})$, which may be read off directly
from (\ref{A1c}),(\ref{A1h}):
\be \boalpha_k^- = \boalpha_k^+ + \bR_k^{-1}\br_k, \lb{53n} \ee
\be {\bf A}_k^- = {\bf A}_k^+ - {{1}\over{2}}\bR_k^{-1}, \lb{53o} \ee
for $k=1,...,n.$ Further details, including the formulation for continuous-time
observation, will be given
elsewhere. We only note here that there is a price to be paid for constructing
a closure of the KSP equation:
the necessity of including among the closure variables the {\it squares} of the
observed variables in addition
to those variables themselves.

\newpage

\section{Conclusion}

This paper is intended to serve as a primer and technical reference for the
application of the proposed
variational estimation scheme to concrete problems. We have discussed the
meaning of the variational estimator
within ensemble theory and emphasized its character as a ``mean-field
approximation'' to the optimal
estimator. Neither the variational method nor the optimal KSP method can be
directly applied in practice
to complex, high-dimensional systems. An action functional can be used to
construct Rayleigh-Ritz
or moment-closure approximations of both the variational and KSP estimators,
but the variational
scheme has the advantage of requiring simpler, lower order closures. We have
discussed a number
of special closure schemes, based in particular upon exponential {\it
Ans\"{a}tze}, that preserve
good properties of the exact estimators. We have discussed also the numerical
implementation of the
variational estimation scheme, both exactly and within a Rayleigh-Ritz
approximation, both to obtain
the estimator itself and also to approximate the variance or ensemble
dispersion. Most of the algorithms
discussed here have already been implemented in \cite{EyRes} and in our
forthcoming work \cite{EyRes1}.

In addition to providing a practical estimation scheme, we hope that the
variational framework will provide
also some additional physical insight into the complex stochastic systems to
which it is applied. It
exploits a thermodynamic formalism for far from equilibrium systems and
provides a motivation to understand
better the concepts of action and entropy in concrete physical systems, e.g.
atmospheres, oceans, ecosystems,
living organisms, etc.

\vspace{.3in}

\noindent {\bf Acknowledgements.} The author wishes to thank F. Alexander, M.
Anitescu, C. E. Leith,
C. D. Levermore and J. Restrepo for valuable conversations and suggestions
which contributed to this work.
He thanks the Isaac Newton Institute for its hospitality during his stay there
for the 1999 Turbulence Programme,
when part of this work was done. This paper was prepared as Los Alamos report
LA-UR00-5264 and supported by the
DOE grant LDRD - ER 2000047.

\newpage

\setcounter{section}{0}
\renewcommand{\thesection}{\Alph{section}}

\section{Appendices}

\noindent {\bf Appendix 1: Optimal Estimation with Discrete-Time Data}

\noindent We give here a simple derivation of the
Kushner-Stratonovich-Pardoux equations for estimation with data
taken at a discrete set of times $t_k,\,\,k=1,...,n$. The problem
set-up is the same as described in Section I.5. We define
$\cP_*(\bx,t):=\cP(\bx,t|\br_1,...,\br_k)$ for $t_{k+1}>t\geq t_k$, so that
$\cP_*(t)$ is right-continuous in time. It is then clear that between
measurement times, $\cP_*(t)$ evolves by the forward Kolmogorov equation
(\ref{6n}).
At measurement times,
\be \cP_*(\bx,t_k+)= \cP_*(\bx,t_k-|\bZ(t_k)+\borho_k=\br_k) \lb{A1a} \ee
for $k=1,...,n$. Thus, by Bayes' rule,
\be
\cP_*(\bx,t_k+)={{\cP_*(\bZ(t_k)+\borho_k=\br_k|\bx,t_k-)\cP_*(\bx,t_k-)}\over
                  {\int
d\by\,\,\cP_*(\bZ(t_k)+\borho_k=\br_k|\by,t_k-)\cP_*(\by,t_k-)}}. \lb{A1b} \ee
By our assumptions, $\borho_k$ is a normal random variable of mean
$\bzed$ and covariance $\bR_k,$ independent of the process
$\bX(t)$. Hence, if $\bZ,\,\borho$ are $s$-dimensional
\be \cP_*(\bZ(t_k)+\borho_k=\br_k|\bx,t_k-)= {{1}\over{\sqrt{(2\pi)^s{\rm
Det}\,\bR_k}}}
\exp\left[-{{1}\over{2}}\left(\bcZ(\bx,t_k)
-\br_k\right)^\top\bR_k^{-1}\left(\bcZ(\bx,t_k)-\br_k\right)\right].
    \lb{A1c} \ee
The term ${{1}\over{\sqrt{(2\pi)^s{\rm
Det}\,\bR_k}}}\exp\left[-{{1}\over{2}}\br_k^\top\bR_k^{-1}\br_k\right]$
may be cancelled between numerator and denominator in (\ref{A1b}). Hence we
obtain finally the forward ``jump condition''
\be \cP_*(\bx,t_k+)=
{{\exp\left[\br_k^\top\bR_k^{-1}\bcZ(\bx,t_k)
-{{1}\over{2}}\bcZ^\top(\bx,t_k)\bR_k^{-1}\bcZ(\bx,t_k)\right]}
                  \over{\cW(\br_1,...,\br_k)}}\cP_*(\bx,t_k-) \lb{A1d} \ee
with the normalization factor
\be \cW(\br_1,...,\br_k):= \int d\by\,\,
\exp\left[\br_k^\top\bR_k^{-1}\bcZ(\by,t_k)-{{1}\over{2}}\bcZ^\top(\by,t_k)
\bR_k^{-1}\bcZ(\by,t_k)\right]\cP_*(\by,t_k-). \lb{A1e} \ee
Next, we {\it define} for $t_{k-1}<t\leq t_k$,
\be \cA_*(\bx,t):= {{\cP(\bx,t|\br_1,...,\br_n)}\over{\cP_*(\bx,t)}}
                 =
{{\cP(\bx,t|\br_1,...,\br_n)}\over{\cP(\bx,t|\br_1,...,\br_{k-1})}}  \lb{A1f}
\ee
and for $t>t_n$.
\be \cA_*(\bx,t):=1 \lb{A1fa} \ee
Writing this definition as
\be \cA_*(\bx,t):=
{{\cP(\bx,t|\br_1,...,\br_n)}\over{\cP(\bx,t|\br_1,...,\br_k)}}
                   \cdot
{{\cP(\bx,t|\br_1,...,\br_k)}\over{\cP(\bx,t|\br_1,...,\br_{k-1})}}
    \lb{A1g} \ee
and using the already derived condition (\ref{A1d}), we obtain for
$t\rightarrow t_k-$ that
\be \cA_*(\bx,t_k-)=\cA_*(\bx,t_k+){{\exp\left[\br_k^\top\bR_k^{-1}
\bcZ(\bx,t_k)-{{1}\over{2}}\bcZ^\top(\bx,t_k)\bR_k^{-1}\bcZ(\bx,t_k)\right]}
                    \over{\cW(\br_1,...,\br_k)}}. \lb{A1h} \ee
This is the backward ``jump condition''.

It remains only to show that $\cA_*(\bx,t)$ defined via (\ref{A1f})
satisfies the backward Kolmogorov equation (\ref{6o}) between measurements.
We apply again Bayes' rule, in the form
\be \cP(\bx,t|\br_1,...,\br_n)=
{{\cP(\br_k,...,\br_n|\bx,t;\br_1,...,\br_{k-1})\cP(\bx,t|\br_1,...,\br_{k-1})}
\over{\cP(\br_k,...,\br_n|\br_1,...,\br_{k-1})}}. \lb{A1i} \ee
However, by the Markov property,
\begin{eqnarray}
\cP(\br_k,...,\br_n|\bx,t;\br_1,...,\br_{k-1}) & = & \cP(\br_k,...,\br_n|\bx,t)
\cr
                                           \,  & = & \int
d\by_k\,\,\cP(\br_k,...,\br_n|\by_k,t_k)\cP(\by_k,t_k|\bx,t)
                                           \lb{A1j} \end{eqnarray}
when $t_{k-1}<t\leq t_k$. Putting together (\ref{A1f}),(\ref{A1i}),(\ref{A1j}),
we conclude that
\be \cA_*(\bx,t)= \int
d\by_k\,\,{{\cP(\br_k,...,\br_n|\by_k,t_k)}
\over{\cP(\br_k,...,\br_n|\br_1,...,\br_{k-1})}}
                                \cP(\by_k,t_k|\bx,t). \lb{A1k} \ee
Since the transition probability satisfies the backward equation in the
variables $\bx,t$
\be (\partial_t + \hL^*)\cP(\by_k,t_k|\bx,t)=0, \lb{A1l} \ee
it then immediately follows from the integral representation (\ref{A1k}) that
$\cA_*(t)$ satisfies (\ref{6o})
for $t_{k-1}<t<t_k,\,\, k=1,...,n.$

It is not hard to show that the jump conditions above, (\ref{A1d}),(\ref{A1h}),
are equivalent to those given in the text,
(\ref{20z1}),(\ref{20z2}). One simply multiplies the numerators and
denominators in (\ref{A1d}),(\ref{A1h}) by the factor
$\exp\left[-{{1}\over{2}}\langle\bZ(t_k)\rangle_{t_k-}^\top\bR_k^{-1}
\langle\bZ(t_k)\rangle_{t_k-}\right]$ and rearranges terms
in the exponents by completing the square.

\newpage

This is an appropriate place to discuss the close formal resemblance of the KSP
jump conditions, (\ref{20z1}),(\ref{20z2}),
to the jump conditions, (\ref{20k}),(\ref{20l}), employed in calculating the
multitime entropy $H_Z$ (or, more correctly,
its Legendre dual $F_Z$.) In fact, the exponential PDF {\it Ansatz} (\ref{38})
has also an interpretation as a conditional PDF.
The conditioning is now upon the event that the empirical average
$\overline{\bZ}_N=\bz$ in the limit as $N\rightarrow\infty$:
\be \lim_{N\rightarrow\infty}\cP_*(\bx,t|\overline{\bZ}_N(t)=\bz)
={{\exp(\bolambda^\top\bcZ(\bx,t))}\over{\cN(\bolambda,t)}}\cP_*(\bx,t)
\lb{A1m} \ee
with $\bolambda=\bolambda(\bz,t)$. More precisely, the result is that
$\lim_{N\rightarrow \infty}\cP_*^{\otimes N}
(\bx_1,..,\bx_N,t|\overline{\bZ}_N(t)=\bz)= \prod_{i=1}^N
{{\exp(\bolambda^\top\bcZ(\bx_i,t))}\over{\cN(\bolambda,t)}}\cP_*(\bx_i,t).$
Here the product measure $\cP_*^{\otimes N}(\bx_1,..,\bx_N;t)=\prod_{i=1}^N
\cP_*(\bx_i,t)$ is taken, to correspond to an
ensemble of $N$ independently prepared samples. Convergence to the new product
measure holds for any finite-dimensional
marginals (i.e. for $i\in S,$ any finite set, as $N\rightarrow\infty$).
Statistical physicists will recognize this as an
equivalence of ensembles result, in which the ``microcanonical ensemble''
corresponding to the condition $\overline{\bZ}_N(t)=\bz$
becomes equivalent in the thermodynamic limit to the ``canonical ensemble''
with potential $\bolambda(\bz,t)$.

As a consequence of this, we may interpret the solution $\cP_*(\bx,t)$ of the
forward equation, with the jump conditions (\ref{20k})
at measurement times less than $t,$ as
\be \cP_*(\bx,t) = \cP(\bx,t|\bz_1,...,\bz_k),\,\,\,\,\,\,\,\,\,t_{k+1}>t\geq
t_k, \lb{A1n} \ee
where the righthand side is shorthand for the PDF conditioned upon the event
$\overline{\bZ}_N(t_1)=\bz_1,...,$
$\overline{\bZ}_N(t_k)=\bz_k$ in the limit $N\rightarrow\infty$. Likewise,
\be \cA_*(\bx,t)\cP_*(\bx,t) = \cP(\bx,t|\bz_1,...,\bz_n), \lb{A1o} \ee
where the conditioning is now upon $\overline{\bZ}_N(t_i)=\bz_i$ for the full
set of times $t_i,\,\,i=1,...,n$, both those before
and after the time $t$. The proof of this assertion is exactly the same as for
the corresponding results proved earlier in this
appendix regarding PDF's conditioned upon observations $\br_1,...,\br_n$. This
relation to conditional PDF's helps to explain
the close similarity to the KSP formalism. Note, however, that this is the
wrong set of conditions to use for estimation, as it
is upon $\overline{\bZ}_N$ itself and not upon (empirical means of)
observations $\overline{\br}_N=\overline{\bZ}_N + \overline{\borho}_N$.

\newpage

\noindent {\bf Appendix 2: Adjoint Calculation of a Derivative}

\noindent For completeness, we shall give a direct proof of (\ref{53a}) here.
We first observe by (\ref{20v2}) and causality that
\be {{\partial F}\over{\partial\bolambda_k}} = \sum_{l=k}^n{{\partial(\Delta
F)_l}\over{\partial\bolambda_k}}. \lb{54} \ee
Then we note from (\ref{52}) that
\be {{\partial(\Delta F)_k}\over{\partial\bolambda_k}}=\bomu_k^+ \lb{55} \ee
while
\be {{\partial(\Delta F)_k}\over{\partial\bolambda_k^-}}=\bomu_k^+-\bomu_k^-.
\lb{56} \ee
Thus, by (\ref{54}),(\ref{55}) and the chain rule
\be {{\partial F}\over{\partial\bolambda_k}}= \bomu_k^+
+\sum_{l=k+1}^n{{\partial(\Delta F)_l}\over{\partial\bomu_l^-}}
         {{\partial\bomu_l^-}\over{\partial\bolambda_k}}. \lb{57} \ee
Furthermore, by (\ref{56}) and the chain rule,
\begin{eqnarray}
{{\partial(\Delta F)_l}\over{\partial\bomu_l^-}} & = & {{\partial(\Delta
F)_l}\over{\partial\bolambda_l^-}}
{{\partial\bolambda_l^-}\over{\partial\bomu_l^-}} \cr
                                             \, & = &
[\bomu_l^+-\bomu_l^-]^\top\boGamma_l^-.  \lb{58}
\end{eqnarray}

Therefore, it only remains in (\ref{57}) to evaluate
${{\partial\bomu_l^-}\over{\partial\bolambda_k}}$ for $l>k$.
The Jacobian matrix for arbitrary times $t\neq t_l,\,l>k$ satisfies the
linearized equation
\be \partial_t {{\partial\bomu(t)}\over{\partial\bolambda_k}}=
\bA(t){{\partial\bomu(t)}\over{\partial\bolambda_k}}, \lb{59} \ee
where $\bA(t):={{\partial\bV}\over{\partial\bomu}}(\bomu(t),t)$. The initial
condition
\be \left.{{\partial\bomu(t)}\over{\partial\bolambda_k}}\right|_{t=t_k+} =
\bC_k^+ \lb{60} \ee
is provided by the formula $\bomu_k^+=\bomu(\bolambda_k^-+\bolambda_k)$ whence
${{\partial\bomu_k^+}\over{\partial\bolambda_k}}=\bC_k^+.$
At the measurement times $t=t_l,\,\,l>k$ there is an additional multiplicative
factor, which follows from the Jacobian
\be {{\partial\bomu^+_l}\over{\partial\bomu_l^-}} =
{{\partial\bomu^+_l}\over{\partial\bolambda_l^-}}
    {{\partial\bolambda^-_l}\over{\partial\bomu_l^-}} = \bC_l^+\boGamma_l^-.
\lb{61} \ee
The solution for $t_l>t>t_{l-1},\,\,l>k$, is
\be {{\partial\bomu(t)}\over{\partial\bolambda_k}}=
    {\rm T}\exp\left[\int_{t_{l-1}}^t \bA(s)\,\,ds\right]
    \left\{ \prod^{l-1}_{j=k+1} \bC_j^+\boGamma_j^-\cdot
    {\rm T}\exp\left[\int_{t_{j-1}}^{t_j} \bA(s)\,\,ds\right]
    \right\}\bC_k^+. \lb{62} \ee
Here ${\rm T}\exp$ denotes the time-ordered exponential with matrices at
increasing times to the left,
and likewise $\Pi$ is the time-ordered product in the same sense. Thus, the
final result
\be {{\partial\bomu_l^-}\over{\partial\bolambda_k}}=
    {\rm T}\exp\left[\int_{t_{l-1}}^{t_l} \bA(s)\,\,ds\right]
    \left\{ \prod^{l-1}_{j=k+1} \bC_j^+\boGamma_j^-\cdot
    {\rm T}\exp\left[\int_{t_{j-1}}^{t_j} \bA(s)\,\,ds\right]
    \right\}\bC_k^+ \lb{63} \ee
follows upon setting $t=t_l-$.

This may be compared with the solution of the adjoint equation
\be \partial_t\boalpha(t)+\bA^*(t)\boalpha(t)=\bzed \lb{64} \ee
integrated backward in time for $t\neq t_l$ and subject to the jump conditions
(\ref{51}) at $t=t_l,\,\,l=1,...,n$.
The explicit solution for $t_k<t<t_{k+1}$ is
\be \boalpha(t)= \sum_{l=k+1}^n
    \overline{{\rm T}}\exp\left[\int^{t_{k+1}}_t \bA^*(s)\,\,ds\right]
    \left\{ \overline{\prod}^{k+2}_{j=l} \boGamma_{j-1}^-\bC_{j-1}^+\cdot
    \overline{{\rm T}}\exp\left[\int_{t_{j-1}}^{t_j} \bA^*(s)\,\,ds\right]
    \right\}\boGamma_l^-[\bomu_l^+-\bomu_l^-]. \lb{65} \ee
Now $\overline{{\rm T}}\exp$ denotes anti-time-ordered exponential with
matrices at decreasing times to the left,
and $\overline{\Pi}$ is the anti-time-ordered product. Setting $t=t_k+$ and
regrouping terms gives
\be \boalpha_k^+= \sum_{l=k+1}^n
    \left\{ \overline{\prod}^{k+1}_{j=l-1} \overline{{\rm
T}}\exp\left[\int^{t_{j-1}}_{t_j} \bA^*(s)\,\,ds\right]
    \cdot \boGamma_j^-\bC_j^+\right\}\overline{{\rm
T}}\exp\left[\int_{t_{l-1}}^{t_l} \bA^*(s)\,\,ds\right]
    \boGamma_l^-[\bomu_l^+-\bomu_l^-]. \lb{66} \ee
Finally, substituting (\ref{58}),(\ref{63}) into (\ref{57}), and using
(\ref{66}), gives
\be {{\partial F}\over{\partial\bolambda_k}} =\bomu_k^+
+\left(\boalpha_k^+\right)^\top\bC_k^+  = \bm_k, \lb{67} \ee
which is exactly the result required.


\newpage

\begin{thebibliography}{99}
\bibitem[1]{Gelb} A. Gelb, ed., {\it Applied Optimal Estimation}. (MIT Press,
Cambridge, MA, 1974).
\bibitem[2] {Strat} R. L. Stratonovich, ``Conditional Markov processes,''
Theor. Prob. Appl.
{\bf 5}: 156-178 (1960).
\bibitem[3] {Kush1} H. J. Kushner, ``On the differential equations satisfied by
conditional probability
densities of Markov processes, with applications,'' J. SIAM Control, Ser.A
{\bf 2}: 106-119 (1962).
\bibitem[4] {Kush2} H. J. Kushner, ``Dynamical equations for optimal nonlinear
filtering,'' J. Diff. Eq.
{\bf 3}: 179-190 (1967).
\bibitem[5]{Kush4} H. J. Kushner, {\it Probability Methods for Approximations
in Stochastic Control and
for Elliptic Equations}. (Academic Press, New York, 1977).
\bibitem[6]{Pard} E. Pardoux, ``\'{E}quations du filtrage non lin\'{e}aire de
la pr\'{e}diction et du lissage,''
Stochastics {\bf 6}: 193-231 (1982).
\bibitem[7]{KalBuc} R. E. Kalman and R. S. Bucy, ``New results in linear
filtering and prediction theory,''
J. Basic Eng. {\bf 83}, Ser.D: 95-108 (1961).
\bibitem[8]{Med} J. S. Meditch, ``On state estimation for distributed parameter
systems,'' J. Franklin Inst.
{\bf 290}: 49-59 (1970).
\bibitem[9]{Kush3}H. J. Kushner, ``Approximation to optimal nonlinear
filters,'' IEEE Trans. Auto.
Contr. {\bf 12} 546-556 (1967).
\bibitem[10]{BHL}D. Brigo, B. Hanzon, and F. LeGland, ``A differential
geometric approach to nonlinear
filtering: the projection filter,'' IEEE Trans. Auto. Contr. {\bf 43} 247-252
(1998).
\bibitem[11]{Ey1}G. L. Eyink, ``Action principle in nonequilibrium statistical
dynamics,'' Phys. Rev. E
{\bf 54}: 3419-3435 (1996).
\bibitem[12]{Ey2}G. L. Eyink, ``Linear stochastic models of nonlinear dynamical
systems,'' Phys. Rev. E
{\bf 58}: 6975-6991 (1998).
\bibitem[13]{Ey3}G. L. Eyink, ``Fluctuation-response relations for multi-time
correlations,''
Phys. Rev. E {\bf 62}: 210-220 (2000).
\bibitem[14]{OM} L. Onsager and S. Machlup, ``Fluctuations and irreversible
processes,'' Phys. Rev. {\bf 91}:
1505-1512 (1953).
\bibitem[15]{Gr}R. Graham, ``Path integral methods in nonequilibrium
thermodynamics and statistics,'' in:
{\it Stochastic Processes in Nonequilibrium Systems}, L. Garrido, P. Seglar,
and P. J. Shepherd, eds.,
Lecture Notes in Physics, vol.84 (Springer-Verlag, Berlin, 1978).
\bibitem[16]{FW}M. I. Freidlin and A. D. Wentzell, {\it Random Perturbations of
Dynamical Systems.}
(Springer-Verlag, New York, 1984).
\bibitem[17]{IZ}C. Itzykson and J.-B. Zuber, {\it Quantum Field Theory} (McGraw
Hill, New York, 1985).
\bibitem[18]{Cramer} H. Cram\'{e}r, ``Sur un noveaux theor\`{e}me-limite de la
theorie des
probabilit\'{e}s,'' Actualit\'{e}s Scientifiques et Industrielles, {\bf 736}:
5-23 (1938).
\bibitem[19]{BZ}R. R. Bhahadur and S. L. Zabell, ``Large deviations of the
sample mean in general
vector spaces,'' Ann. Prob. {\bf 7} 587-621 (1979).
\bibitem[20]{V}S. R. S. Varadhan, {\it Large Deviations and Applications}.
(SIAM, Philadelphia, 1984).
\bibitem[21]{Leith}C. E. Leith, ``Theoretical skill of Monte Carlo forecasts,''
Mon. Wea. Rev.
{\bf 102} 409-418 (1974).
\bibitem[22]{BLE}G. Burgers, P. J. van Leeuwen, and G. Evensen, ``Analysis
schemes in the ensemble Kalman
filter,'' Mon. Wea. Rev. {\bf 126} 1719-1724 (1998).
\bibitem[23] {NocWr} J. Nocedal and S. J. Wright, {\it Numerical Optimization.}
Springer Series in Optimization
             Research. (Springer, New York, 1999).
\bibitem[24]{KGV} S. Kirkpatrick, C. D. Gelatt, and M. P. Vecchi,
``Optimization by simulated
annealing,'' Science {\bf 220}: 671-680 (1983).
\bibitem[25]{Kirk} S. Kirkpatrick, ``Optimization by simulated annealing:
quantitative studies,''
J. Stat. Phys. {\bf 34}: 975-986 (1984).
\bibitem[26]{Zak}M. Zakai, ``On the optimal filtering of diffusion
  processes,'' Z. Wahrscheinlichkeitstheorie verw. Geb. {\bf 11} 230-243
(1969).
\bibitem[27]{GL}G. Eyink and C. D. Levermore, ``Entropy-based closure of
nonlinear stochastic dynamics,''
                in preparation.
\bibitem[28]{CDL}C. D. Levermore, ``Moment closure hierarchies for kinetic
theories,'' J. Stat. Phys.
                 {\bf 83} 1021-1065 (1996).
\bibitem[29]{Ey4} G. Eyink, ``Rayleigh-Ritz effective action in a double
exponential Ansatz,''
                  in preparation.
\bibitem[30]{EyRes} G. Eyink and J. M. Restrepo, ``Most probable histories for
nonlinear dynamics:
                    tracking climate transitions,'' J. Stat. Phys. {\bf 101}
459-472 (2000).
\bibitem[31]{EyRes1} G. Eyink and J. M. Restrepo, ``Optimal variational
assimilation in strongly nonlinear dynamical
                    systems,'' in preparation.

\end{thebibliography}
\end{document}